\documentclass[preprintnumbers,article,amsmath,amssymb,floatfix,10pt,prd,onecolumn,
superscriptaddress,nofootinbib]{revtex4}
\usepackage{bm}
\usepackage{amsfonts}
\usepackage{latexsym}
\usepackage[latin1]{inputenc}
\usepackage{graphicx}
\usepackage{amsmath}
\usepackage{palatino}
\usepackage{mathpazo}
\usepackage{textcomp}
\usepackage{float}
\usepackage{booktabs}
\usepackage{dcolumn}
\usepackage{ragged2e}
\usepackage{hyperref}
\hypersetup{colorlinks,citecolor=blue}
\hypersetup{colorlinks=true,linkcolor=red,filecolor=magenta,    urlcolor=blue}
\usepackage{amsmath}
\usepackage{xcolor}
\usepackage{orcidlink}
\usepackage{epsfig}
\usepackage[caption=false]{subfig}
\usepackage{commath}
\def\jnl@style{\it}
\def\aaref@jnl#1{{\jnl@style#1}}

\def\aaref@jnl#1{{\jnl@style#1}}

\def\aj{\aaref@jnl{AJ}}                   % Astronomical Journal
\def\apj{\aaref@jnl{ApJ}}                 % Astrophysical Journal
\def\apjl{\aaref@jnl{ApJ}}                % Astrophysical Journal, Letters
\def\apjs{\aaref@jnl{ApJS}}               % Astrophysical Journal, Supplement
\def\apss{\aaref@jnl{Ap\&SS}}             % Astrophysics and Space Science
\def\aap{\aaref@jnl{A\&A}}                % Astronomy and Astrophysics
\def\aapr{\aaref@jnl{A\&A~Rev.}}          % Astronomy and Astrophysics Reviews
\def\aaps{\aaref@jnl{A\&AS}}              % Astronomy and Astrophysics, Supplement
\def\mnras{\aaref@jnl{Mon.~Not.~Roy.~Astron.~Soc.}}             % Monthly Notices of the RAS
\def\prd{\aaref@jnl{Phys.~Rev.~D}}        % Physical Review D
\def\prc{\aaref@jnl{Phys.~Rev.~C}}  % Physical Review C
\def\prl{\aaref@jnl{Phys.~Rev.~Lett.}}    % Physical Review Letters
\def\qjras{\aaref@jnl{QJRAS}}             % Quarterly Journal of the RAS
\def\skytel{\aaref@jnl{S\&T}}             % Sky and Telescope
\def\ssr{\aaref@jnl{Space~Sci.~Rev.}}     % Space Science Reviews
\def\zap{\aaref@jnl{ZAp}}                 % Zeitschrift fuer Astrophysik
\def\nat{\aaref@jnl{Nature}}              % Nature
\def\aplett{\aaref@jnl{Astrophys.~Lett.}} % Astrophysics Letters
\def\apspr{\aaref@jnl{Astrophys.~Space~Phys.~Res.}} % Astrophysics Space Physics Research
\def\physrep{\aaref@jnl{Phys.~Rep.}}      % Physics Reports
\def\physscr{\aaref@jnl{Phys.~Scr}}       % Physica Scripta
\def\commat{\aaref@jnl{Comm.~Math.~Phys.}}              % Communications in Mathematical Physics
\def\science{\aaref@jnl{Science}}               % Science
\def\cqg{\aaref@jnl{Classical Quant.~Grav.}}            % Classical and Quantum Gravity
\def\jpcs{\aaref@jnl{JPCS}}                                     % Journal of Physics Conference Series
\def\ijmpd{\aaref@jnl{Int.~J.~Mod.~Phys.~D}}                    % International Journal of Modern Physics D
\def\grg{\aaref@jnl{Gen.~Relat.~Gravit.}}               % General Relativity and Gravitation
\def\rpp{\aaref@jnl{Rep.~Prog.~Phys.}}          % Reports on Progress in Physics
\def\npa{\aaref@jnl{Nucl.~Phys.~A}}        % Nuclear Physics A
\def\lrr{\aaref@jnl{Living Rev.~Rel.}}                   % Living reviews in relativity
\def\jcap{\aaref@jnl{J.~Cosmology Astropart.~Phys.}}    % Journal of cosmology and astroparticle physics
\def\rmp{\aaref@jnl{Rev.~Mod.~Phys.}}   %Reviews of modern physics
\def\epjc{\aaref@jnl{Eur.~Phys.~J.~C}}
\def\plb{\aaref@jnl{~Phy.~Lett.~B}}
\def\mpla{\aaref@jnl{Mod.~Phy.~Lett.~A}}
\def\arxiv{\aaref@jnl{arxiv.org}}

\allowdisplaybreaks[1]
\begin{document}
%\color{red}
\color{black}       %% For one column
\title{Wormhole solutions in symmetric teleparallel gravity}
%\end{document}
\author{G. Mustafa}
\email{gmustafa3828@gmail.com}
\affiliation{Department of Mathematics, Shanghai University,
Shanghai, 200444, Shanghai, People's Republic of China.}

\author{Zinnat Hassan\orcidlink{0000-0002-6608-2075}}
\email{zinnathassan980@gmail.com}
\affiliation{Department of Mathematics, Birla Institute of Technology and
Science-Pilani,\\ Hyderabad Campus, Hyderabad-500078, India.}

\author{P.H.R.S. Moraes}
\email{moraes.phrs@gmail.com}
\affiliation{Universidade de S\~ao Paulo (USP), Instituto de Astronomia, Geof\'isica e Ci\^encias Atmosf\'ericas (IAG), Rua do Mat\~ao 1226, Cidade Universit\'aria, 05508-090 S\~ao Paulo, SP, Brazil}
\affiliation{Instituto Tecnol\'ogico de Aeron\'autica (ITA), Departamento de F\'isica, Centro T\'ecnico Aeroespacial, 12228-900 S\~ao Jos\'e dos Campos, S\~ao Paulo, Brazil}

\author{P.K. Sahoo\orcidlink{0000-0003-2130-8832}}
\email{pksahoo@hyderabad.bits-pilani.ac.in}
\affiliation{Department of Mathematics, Birla Institute of Technology and
Science-Pilani,\\ Hyderabad Campus, Hyderabad-500078, India.}

\date{\today}
\begin{abstract}

In this letter we obtain wormhole solutions from the Karmarkar condition in $f(Q)$ gravity formalism, in which $Q$ is the nonmetricity scalar. We show that the combination of such ingredients provides us the possibility of obtaining traversable wormholes satisfying the energy conditions. Our results contribute to $f(Q)$ gravity ascension, that was recently ignited by the natural prediction of early and late-time stages of universe expansion acceleration in the formalism.

\end{abstract}

\maketitle

\date{\today}

\section{Introduction}

Wormholes and black holes are fascinating solutions of General Relativity (GR). Black holes existence has already been probed \cite{abbott/2016,abbott/2018,Akiyama}. On the other hand, the existence of wormholes in the universe is still an open question. A deep discussion on this possibility was presented in \cite{khatsymovsky/1994}. The possible existence of wormholes in the galactic halo region was discussed in \cite{rahaman/2014,ovgun/2016}. Lensing by wormholes was discussed in \cite{safonova/2002,javed/2019,ovgun/2019} {\bf and light deflection in \cite{ovgun/2018,ovgun/2020}}. Recently, a review of past and current efforts to search for wormholes in the universe was written \cite{bambi/2021}.

Wormholes are tube-like geometrical structures connecting two distinct universes or two different points in the same universe. The wormhole concept was firstly proposed by Flamm \cite{L. Flamm}, who built up the Schwarzschild solution of isometric embedding. After Flamm, Einstein and Rosen \cite{N. Rosen} obtained a comparable geometrical construction which is known as the Einstein-Rosen bridge. In 1957, Wheeler and Misner \cite{Wheeler,Misner} first introduced the term $wormhole$.

Morris and Thorne \cite{Morris} first gave the idea of a traversable wormhole. Using GR, they investigated spherically symmetric static objects and showed that they should violate energy conditions. This violating energy condition exotic matter has physical properties that would violate known laws of physics, such as a particle having negative mass. {\bf In view of these outstanding properties, wormholes are a very interesting topic in theoretical physics \cite{maeda/2021,jusufi/2020,richarte/2017,halilsoy/2014}.}

It has been shown that modified theories of gravity could be able to considerably minimize or even nullify the need for exotic matter. The extra degrees of freedom of such theories make their field equations to present some new terms that can allow the existence of wormholes respecting the energy conditions. Mazharimousavi and Halilsoy have constructed traversable wormholes in the $f(R)$ gravity, for which $R$ stands for the Ricci scalar, and showed that they respect at least the weak energy condition \cite{mazharimousavi/2016}. DeBenedictis and Horvat have also constructed $f(R)$ wormholes respecting energy conditions \cite{debenedictis/2012}. Dehgani and Mehdizadeh have constructed wormholes in Lovelock gravity in seven dimensions \cite{dehgani/2012}. They showed that for negative second-order and positive third-order Lovelock coefficients, there are thin-shell wormholes respecting the weak energy condition. For further modified gravity wormholes one can check \cite{moraes/2019,golchin/2019,godani/2020,baruah/2019,ovgun/2019b}.

It is worth mentioning that wormholes in modified gravity emerge as a new strand in the realm of gravitational theories applications. In the last years it has been quite common to see modified gravity as alternatives to dark energy \cite{bertschinger/2008,lue/2004,langlois/2019} and dark matter \cite{sanders/2007,cembranos/2016,rinaldi/2017}. Modified gravity has also been applied to stellar astrophysics \cite{silva/2016,pani/2011,sakstein/2015,jain/2011,chang/2011} mainly with the purpose of verifying the possibility of increasing the maximum mass of compact objects, such as neutron stars and white dwarfs, in order to keep in touch with recent observations \cite{freire/2008,freire/2008b,linares/2018,van_kerkwijk/2011,gvaramadze/2019}.

In the present letter we are going to consider wormholes in the $f(Q)$ theory of gravity \cite{Jimenez/2018}. It is well known that GR cannot distinguish between gravitation and inertial effects. However, by resorting to frame fields, the gravitational energy can be defined covariantly in the teleparallel approach \cite{maluf/2013}. The canonical frame is then identified by the absence of curvature and torsion, and the canonical coordinates are identified by the absence of inertial effects. In the $f(Q)$ or symmetric teleparallel gravity, $Q$ is the quadratic nonmetricity scalar \cite{Jimenez/2018}

\begin{equation}\label{i1}
	Q=-\frac{1}{4}Q_{\alpha\beta\mu}Q^{\alpha\beta\mu}+\frac{1}{2}Q_{\alpha\beta\mu}Q^{\beta\mu\alpha}+\frac{1}{4}Q_\alpha Q^\alpha-\frac{1}{2}Q_\alpha\tilde{Q}^\alpha,
\end{equation}
with the independent traces denoted as $Q_\mu=Q_\mu\ ^\alpha\ _\alpha$ and $\tilde{Q}^\mu=Q_\alpha\ ^{\mu\alpha}$. Among the general quadratic combinations, \eqref{i1} is special because in addition to being invariant under local general linear transformations, it is invariant under a translational symmetry that allows to completely remove the connection.

Although the $f(Q)$ gravity was quite recently proposed, some interesting applications of it can already be appreciated. We quote, for instance, some cosmological aspects of $f(Q)$ gravity that were investigated in \cite{jimenez/2020,frusciante/2021,bajardi/2020} and the energy conditions that were constructed in $f(Q)$ gravity in Reference \cite{mandal/2020}. For further $f(Q)$ gravity applications one can check References \cite{lazkoz/2019,barros/2020}.

As it was aforementioned, in the present letter, we are going to consider wormhole solutions in the $f(Q)$ gravity. The geometry of the wormhole space-time will be deeply discussed in Section II. The $f(Q)$ gravity basis will be presented in Section III, along with the wormhole field equations in such a theory. The energy conditions are presented and calculated in Section IV for different particular forms of the $f(Q)$ function. Finally, in Section V, we discuss the physics of our results and present our concluding remarks.

\section{The geometry of traversable wormholes, Karmarkar condition and embedding space-time}

We start with spherically symmetric and static space-time, such as
\begin{equation} \label{1}
ds^{2}=e^{\nu(r)}dt^{2}-e^{\lambda(r)} dr^{2}-r^{2} d\theta^{2}-r^{2}\sin^{2}\theta d\phi^{2} ,
\end{equation}
where the components $\nu(r)$ and $\lambda(r)$ of metric space are radial-only dependent functions.

In the present analysis we will obtain wormhole solutions using Karmarkar condition \cite{Karmarkar/1948} with embedded class-1 space-time. This condition is one of the utmost significant features of the current analysis. The basic construction of the Karmarkar condition depends upon the embedded class-1 solution of Riemannian space. Eisenhart provided an essential and appropriate requirement for the embedded class-1 solution \cite{Eisenhart/1966}, which depends on a symmetric tensor of the second order, $b_{mn}$, and on the Riemann curvature tensor, $\mathcal{R}_{mnpq}$, through

\begin{itemize}
     \item the Gauss equation:
 \end{itemize}
\begin{eqnarray}\label{eqcls1.1}
\mathcal{R}_{mnpq}=2\,\epsilon\,{b_{m\,[p}}{b_{q]n}},
\end{eqnarray}
\begin{itemize}
     \item the Codazzi equation:
 \end{itemize}
\begin{eqnarray}\label{eqcls1.2}
b_{m\left[n;p\right]}-{\Gamma}^q_{\left[n\,p\right]}\,b_{mq}+{{\Gamma}^q_{m}}\,{}_{[n}\,b_{p]q}=0.
\end{eqnarray}
Here $\epsilon=\pm1$ and square brackets represent antisymmetrization, while $b_{mn}$ represent the coefficients of the second differential form. Using Eqs.~(\ref{eqcls1.1}) and ~(\ref{eqcls1.2}), by imposing the above mathematical mechanism, the Karmarkar condition is calculated as
\begin{equation}\label{2}
\mathcal{R}_{2323}\mathcal{R}_{1414}=\mathcal{R}_{1224}\mathcal{R}_{1334}+ \mathcal{R}_{1212}\mathcal{R}_{3434},
\end{equation}
with Pandey and Sharma condition \cite{Pandey/1981}, i.e., $\mathcal{R}_{2323}\neq\mathcal{R}_{1414}\neq0$.

By plugging the appropriate Riemanian tensor components in Equation  (\ref{2}) we obtain the following differential equation
\begin{equation}\label{3}
\frac{\nu'(r) \lambda'(r)}{1-e^{\lambda(r)}}-\left\{\nu'(r) \lambda'(r)+\nu'(r)^2-2 \left[\nu''(r)+\nu'(r)^2\right]\right\}=0,\;\;\;\;\;\;\;e^{\lambda(r)}\neq1,
\end{equation}
with primes indicating radial derivatives. By solving (\ref{3}) we obtain
\begin{equation}\label{4}
e^{\lambda(r)}=1+Ae^{\nu(r)}\nu^{'2}(r),
\end{equation}
where $A$ is an integration constant.

Gupta $et\, al.$ \cite{Gupta/1975} have provided a six-dimensional embedded Euclidean space-time,  defined below:
\begin{equation}\label{Eq2.37}
 d{s}^2= dX_{1}^{2}+dX_{2}^{2}\pm dX_{3}^{2}-dx_{1}^{2}-dx_{2}^{2}-dx_{3}^{2},
\end{equation}
where
\begin{eqnarray}
x_{1}&&=r sin \theta cos \phi,\;\;\;\;\; \;\;\;\;\;\;\;\;\;x_{2}=r sin \theta sin \phi,\;\;\;\;\;\;\;\;\;\;\;\;\;\;\;x_{3}=r cos\;\theta,\nonumber\\
X_{1}&&=A e^{\lambda(r)/2} cosh(t/A), \;\;\;X_{2}=A e^{\nu(r)/2} sinh(t/A),\;\;\;X_{3}=H(r).\label{Eq2.38}
\end{eqnarray}
Here, $A>0$ and $H^{'^{2}}(r)=\mp\left\{-[e^{\lambda(r)}-1]+\frac{A^{2}e^{\nu(r)}\nu^{'^{2}}(r)}{4}\right\}$. Using transformation (\ref{Eq2.38}) with space-time (\ref{1}), we get an embedded class one spacetime, which satisfies a Karmarkar condition.

A spherically symmetric space-time for wormhole geometry is defined as:
\begin{equation}\label{5}
ds^{2}=e^{2a(r)}dt^{2}-\frac{1}{1-\frac{b(r)}{r}}dr^{2}-r^{2}d\theta^{2}-r^{2}sin^{2}\theta d\phi^{2}.
\end{equation}

By comparing Equation (\ref{1}) and Equation (\ref{5}), we have the following relations between the gravitational components
\begin{equation}\label{6}
\nu(r)=2a(r),\;\;\;\;\;\;\;\;\;\;\lambda(r)=Log\Big[\frac{r}{r-b(r)}\Big].
\end{equation}

Due to embedded class one solutions, we consider the specific redshift function as \cite{Zia,Anchordoqui}
\begin{equation}\label{7}
\nu(r)=2a(r)=-\frac{2\chi }{r}.
\end{equation}
The chosen redshift function satisfies the flatness condition, i.e., $\chi(r)\rightarrow 0$ when $r \rightarrow \infty$. Moreover, $b(r)$ is known as wormhole shape function. By taking Equation (\ref{4}) and Equation (\ref{6}) into account, we obtain the following relation for the shape function:

\begin{equation}\label{8}
b(r)=r-\frac{r^{5}}{r^{4}+4\zeta^{2}A e^{-\frac{2 \chi }{r}} }.
\end{equation}

According to Morris \cite{Thorne} the calculated shape function of a traversable
wormhole must meet the following properties: (I) $b(r) - r = 0$ at $r = r_{0}$, the wormhole throat; (II) $\frac{b(r)-rb^{'}(r)}{
b(r)} > 0$; (III) $b^{'}(r)<1$ and (IV) $\frac{b(r)}{r}\rightarrow 0$ as $r\rightarrow \infty$, such that the radial coordinate is such that $r_0 < r <\infty $.

By making $b(r_0) - r_0 =0$ we get a trivial solution. To avoid it, we introduce a new free parameter, say $\delta$, and add it into Eq.(\ref{8}). Once again by using condition (I), we get $A=\frac{r_{0}^{4}(r_{0}-\delta)}{4 \chi^{2}e^{\frac{-2\chi}{r}}}$. Finally, Eq. (\ref{8}) takes the following form:
\begin{equation}\label{9}
b(r)=r-\frac{r^{5}}{r^{4}+r_{0}^{4}(r_{0}-\delta)}+\delta,\;\; 0<\delta<r_{0}.
\end{equation}

\section{$f(Q)$ formalism and wormhole field equations}\label{sec2}

The action for symmetric teleparallel gravity is given by \cite{Jimenez/2018}
\begin{equation}
\label{14}
\mathcal{S}=\int\frac{1}{2}\,f(Q)\sqrt{-g}\,d^4x+\int \mathcal{L}_m\,\sqrt{-g}\,d^4x\,,
\end{equation}
where $f(Q)$ represents the function form of Q, $g$ is the determinant of the metric $g_{\mu\nu}$, and $\mathcal{L}_m$ is the matter Lagrangian density.\\

The non-metricity tensor and its traces can be written as\\
\begin{equation}
\label{15}
Q_{\lambda\mu\nu}=\bigtriangledown_{\lambda} g_{\mu\nu}
\end{equation}
\begin{equation}
\label{16}
Q_{\alpha}=Q_{\alpha}\;^{\mu}\;_{\mu},\; \tilde{Q}_\alpha=Q^\mu\;_{\alpha\mu}
\end{equation}
Also, the non-metricity tensor helps us to write the nonmetricity conjugate defined as
\begin{equation}
\label{17}
P^\alpha\;_{\mu\nu}=\frac{1}{4}\left[-Q^\alpha\;_{\mu\nu}+2Q_{(\mu}\;^\alpha\;_{\nu)}+Q^\alpha g_{\mu\nu}-\tilde{Q}^\alpha g_{\mu\nu}-\delta^\alpha_{(\mu}Q_{\nu)}\right]
\end{equation}
One can readily check that the nonmetricity can also be written in the form \cite{Jimenez/2018,jimenez/2020}
\begin{equation}
\label{18}
Q=-Q_{\alpha\mu\nu}\,P^{\alpha\mu\nu},
\end{equation}
where $P^{\alpha\mu\nu}$ is called the nonmetricity conjugate because it satisfies
\begin{equation}
P^{\alpha\mu\nu}=-\frac{1}{2}\frac{\partial Q}{\partial Q_{\alpha\mu\nu}}.
\end{equation}

The energy-momentum tensor for the fluid description of the spacetime can be written as
\begin{equation}
\label{19}
T_{\mu\nu}=-\frac{2}{\sqrt{-g}}\frac{\delta\left(\sqrt{-g}\,\mathcal{L}_m\right)}{\delta g^{\mu\nu}}.
\end{equation}

Now, one can write the motion equations by varying the action \eqref{15} with respect to metric tensor $g_{\mu\nu}$, which can be written as
\begin{equation}
\label{20}
\frac{2}{\sqrt{-g}}\bigtriangledown_\gamma\left(\sqrt{-g}\,f_Q\,P^\gamma\;_{\mu\nu}\right)+\frac{1}{2}g_{\mu\nu}f \\
+f_Q\left(P_{\mu\gamma i}\,Q_\nu\;^{\gamma i}-2\,Q_{\gamma i \mu}\,P^{\gamma i}\;_\nu\right)=-T_{\mu\nu},
\end{equation}
where $f_Q\equiv\frac{df}{dQ}$. Also varying \eqref{15} with respect to the connection,one obtains
\begin{equation}
\label{21}
\bigtriangledown_\mu \bigtriangledown_\nu \left(\sqrt{-g}\,f_Q\,P^\gamma\;_{\mu\nu}\right)=0.
\end{equation}

For the present interest, we consider matter is described by an anisotropic stress-energy tensor of the form
\begin{equation}
\label{22}
T_{\mu}^{\nu}=\left(\rho+p_t\right)u_{\mu}\,u^{\nu}-p_t\,\delta_{\mu}^{\nu}+\left(p_r-p_t\right)v_{\mu}\,v^{\nu}
\end{equation}
where $u_{\mu}$ is the four-velocity, $v_{\mu}$ the unitary space-like vector in the radial direction, $\rho$ is the energy density, $p_r$ is the pressure in the direction of $u_{\mu}$ (radial pressure) and $p_t$ is the pressure orthogonal to $v_{\mu}$(tangential pressure). Here, $p_r$ and $p_t$ are functions of radial component $r$. The trace of the non-metricity tensor $Q$ for the wormhole metric in \eqref{9} takes the form
\begin{equation}
\label{23}
Q=-\frac{2}{r}\left[1-\frac{b(r)}{r}\right]\left[2a^{'}(r)+\frac{1}{r}\right].
\end{equation}
Now, by substituting \eqref{5} and \eqref{22} in \eqref{20}, one can find the following field equations

\begin{equation}
\label{24}
\left[\frac{1}{r}\left(-\frac{1}{r}+\frac{rb^{'}(r)+b(r)}{r^2}-2a^{'}(r)\left(1-\frac{b(r)}{r}\right)\right)\right]f_Q
-\frac{2}{r}\left(1-\frac{b(r)}{r}\right)\dot{f}_Q-\frac{f}{2}=-\rho,
\end{equation}
\begin{equation}
\label{25}
\left[\frac{2}{r}\left(1-\frac{b(r)}{r}\right)\left(2a^{'}(r)+\frac{1}{r}\right)-\frac{1}{r^2}\right]f_Q+\frac{f}{2}=-p_r,
\end{equation}
\begin{multline}
\label{26}
\left[\frac{1}{r}\left(\left(1-\frac{b(r)}{r}\right)\left(\frac{1}{r}+a^{'}(r)\left(3+ra^{'}(r)\right)+ra^{''}(r)\right)
-\frac{rb^{'}(r)-b(r)}{2r^2}\left(1+ra^{'}(r)\right)\right)\right]f_Q+\\
\frac{1}{r}\left(1-\frac{b(r)}{r}\right)\left(1+ra^{'}(r)\right)\dot{f}_Q+\frac{f}{2}=-p_t.
\end{multline}

\section{Wormhole solutions and Energy Conditions}

The Raychaudhuri equations state the temporal evolution of expansion scalar ($\theta$) for the congruences of timelike ($u^\mu$) and null ($\eta_\mu$) geodesics as \cite{Raychaudhuri/1955,S.Nojiri,J.Ehlers}
\begin{equation}
\label{27}
\frac{d\theta}{d\tau}-\omega_{\mu\nu}\,\omega^{\mu\nu}+\sigma_{\mu\nu}\sigma^{\mu\nu}+\frac{1}{3}\theta^2+R_{\mu\nu}u^\mu\,u^\nu=0,
\end{equation}
\begin{equation}
\label{28}
\frac{d\theta}{d\tau}-\omega_{\mu\nu}\,\omega^{\mu\nu}+\sigma_{\mu\nu}\sigma^{\mu\nu}+\frac{1}{2}\theta^2+R_{\mu\nu}\eta^\mu\eta^\nu=0,
\end{equation}
where $\sigma^{\mu\nu}$ and $\omega_{\mu\nu}$ are the shear and the rotation associated with the vector field $u^\mu$ respectively. For attractive nature of gravity ($\theta<0$) and neglecting the quadratic terms, the Raychaudhuri equations \eqref{27} and \eqref{28} satisfy the following conditions
\begin{equation}
\label{29}
R_{\mu\nu}u^\mu\,u^\nu\geq0,
\end{equation}
\begin{equation}
\label{30}
R_{\mu\nu}\eta^\mu\eta^\nu\geq0.
\end{equation}
As we are working with anisotropic fluid matter distribution, the energy condition recovered from GR are\\
$\bullet$ Strong energy conditions (SEC) if $\rho+p_j\geq0$, $\rho+\sum_jp_j\geq0$, $\forall j$.\\
$\bullet$ Dominant energy conditions (DEC) if $\rho\geq0$, $\rho\pm p_j\geq0$, $\forall j$.\\
$\bullet$ Weak energy conditions (WEC) if $\rho\geq0$, $\rho+p_j\geq0$, $\forall j$.\\
$\bullet$ Null energy condition (NEC) if $\rho+p_j\geq0$, $\forall j$.
where $\rho$ and $p$ describe the energy density and pressure, respectively.

\subsection{The $f(Q)=Q-\alpha \gamma \left(1-e^{-\frac{Q}{\gamma}}\right)$ case}

Herein, we propose a specific exponential type model for the $f(Q)$ function, which is expressed as 
\begin{equation}
\label{31}
f(Q)=Q-\alpha \gamma \left(1-e^{-\frac{Q}{\gamma}}\right),
\end{equation}
where $\alpha$ and $\gamma$ are the model free parameters.

By plugging Equations (\ref{7}), (\ref{8}) and (\ref{31}) into Equations (\ref{24})-(\ref{26}), we can write our wormhole material solutions as follows:

\begin{eqnarray}
\rho&=&\frac{1}{2} \left(\alpha  \gamma  \left(\Lambda _1-1\right)+\frac{2}{r^4}\left(\frac{4 \alpha  \Lambda _1 \Lambda _2}{\gamma  r^3}+2 \Lambda _3 \left(\alpha  \Lambda _1-1\right)-(r-b(r)) (4 \chi +r)\right)\right),\label{32}\\
p_{r}&=&\frac{\frac{1}{r}\left(8 \chi  b(r)-\alpha  \Lambda _1 \left(4 b(r) (4 \chi +r)+r \left(-16 \chi +\gamma  r^3-2 r\right)\right)\right)+2 b(r)-8 \chi +\alpha  \gamma  r^3}{2 r^3},\label{33}\\
p_{t}&=&\frac{-\frac{4 \alpha  \Lambda _1 \Lambda _2 (2 \chi +r)}{\gamma  r^2}+r\left(\Lambda _5 r-\Lambda _4 \left(\alpha  \Lambda _1-1\right)\right)}{2 r^6},\label{34}
\end{eqnarray}
where
\begin{eqnarray*}
\Lambda _1 &&\equiv e^{\frac{2 (r-b(r)) (4 \chi +r)}{\gamma  r^4}},\\
\Lambda _2 &&\equiv(r-b(r)) \left(r (4 \chi +r) b'(r)-b(r) (16 \chi +3 r)+2 r (6 \chi +r)\right),\\
\Lambda _3 &&\equiv r^2 b'(r)-(r-b(r)) (4 \chi +r),\\
\Lambda _4 &&\equiv r \left(2 \chi +r\right) \left(r b'(r)-2 \left(2 \chi +r\right)\right)+b(r) \left(8 \chi ^2+r^2+2 \chi  r\right),\\
\Lambda _5 &&\equiv r \left(8 \chi +\alpha  \gamma  \left(-\left(\Lambda_1-1\right)\right) r^3+2 r\right)-2 b(r) (4 \chi +r).
\end{eqnarray*}

In Figures 1-6 we plot the energy conditions for the material solutions written above and taking into account our solution for the shape function. When plotting those we fix the value of the parameter $\gamma$. We discuss such results later in Section V. 

Below we visit a second case for the $f(Q)$ function. Here, it is important to clarify that the functional forms we are using in the $f(Q)$ gravity formalism are motivated by some $f(R)$ forms already present in the literature, as one can check References \cite{Pavlovic/2015,Elizalde/2011,Tsujikawa/2008}.

\begin{figure}
\centering \epsfig{file=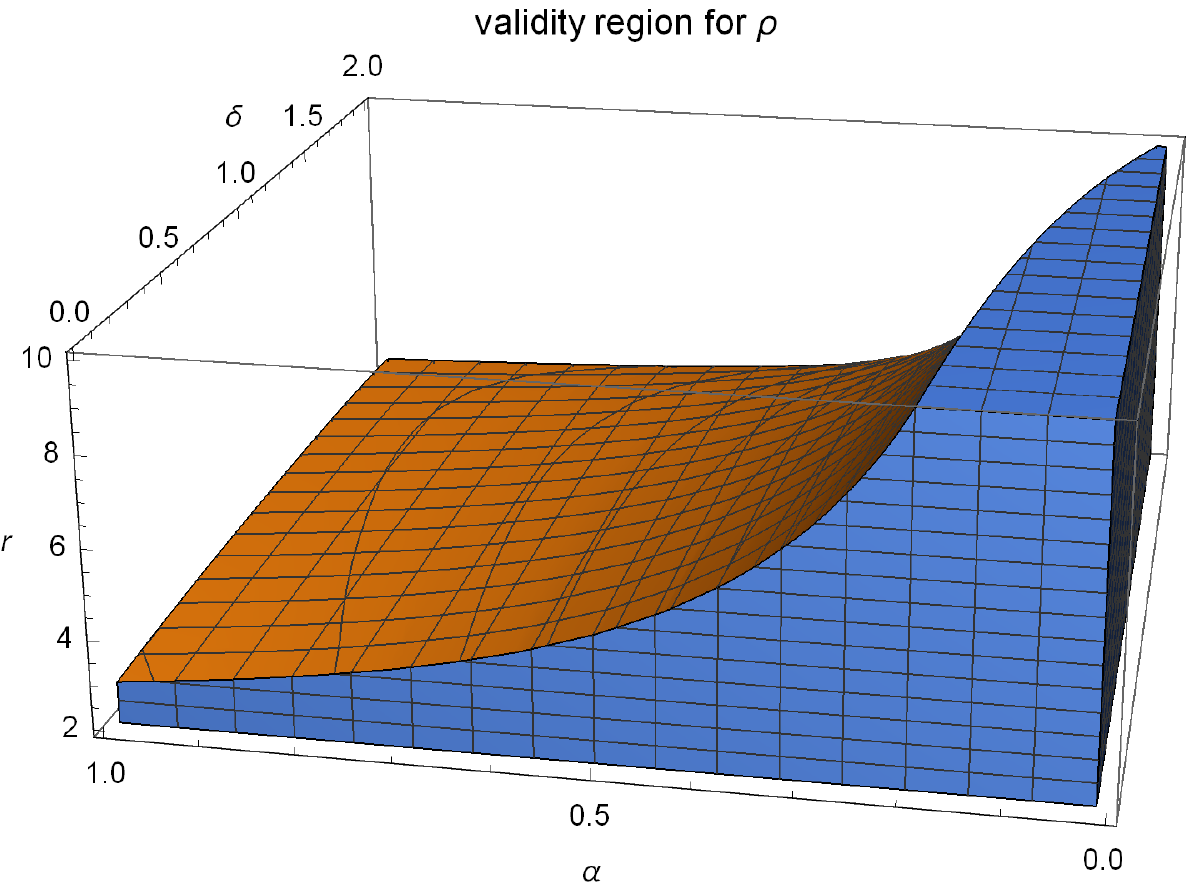, width=.45\linewidth,
height=2.5in}\epsfig{file=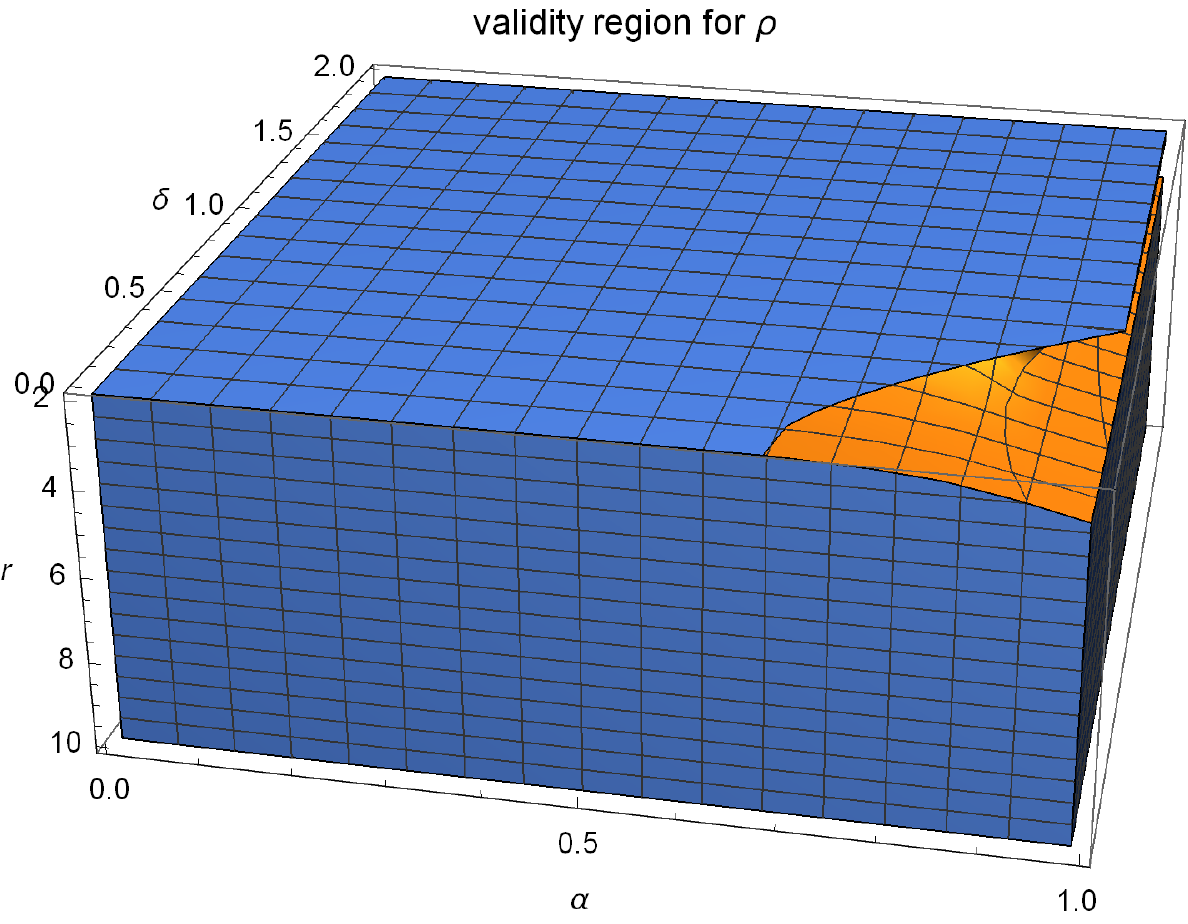, width=.45\linewidth,
height=2.5in}\caption{\label{Fig1} Energy density $\rho(r)$ with $\gamma=-2$ (left) and $\gamma=6$ (right) for different values of $\alpha$ and $\delta$.}
\end{figure}
\begin{figure}
\centering \epsfig{file=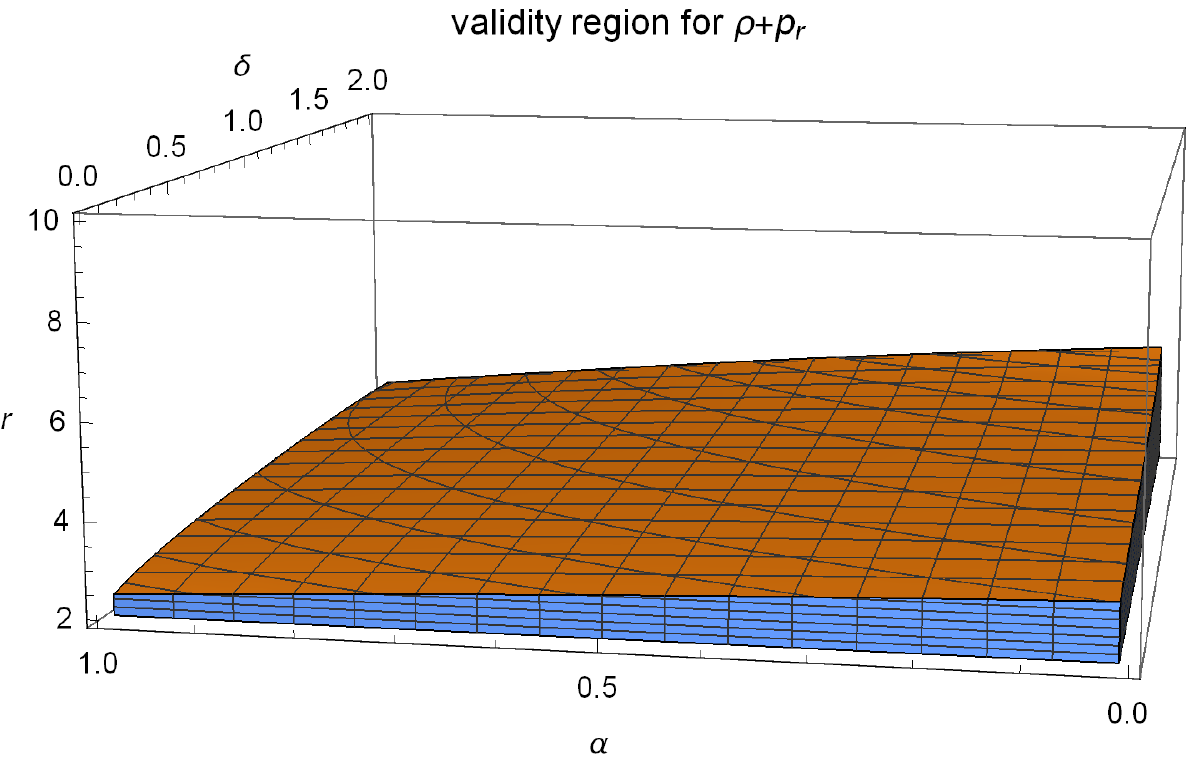, width=.45\linewidth,
height=2.5in}\epsfig{file=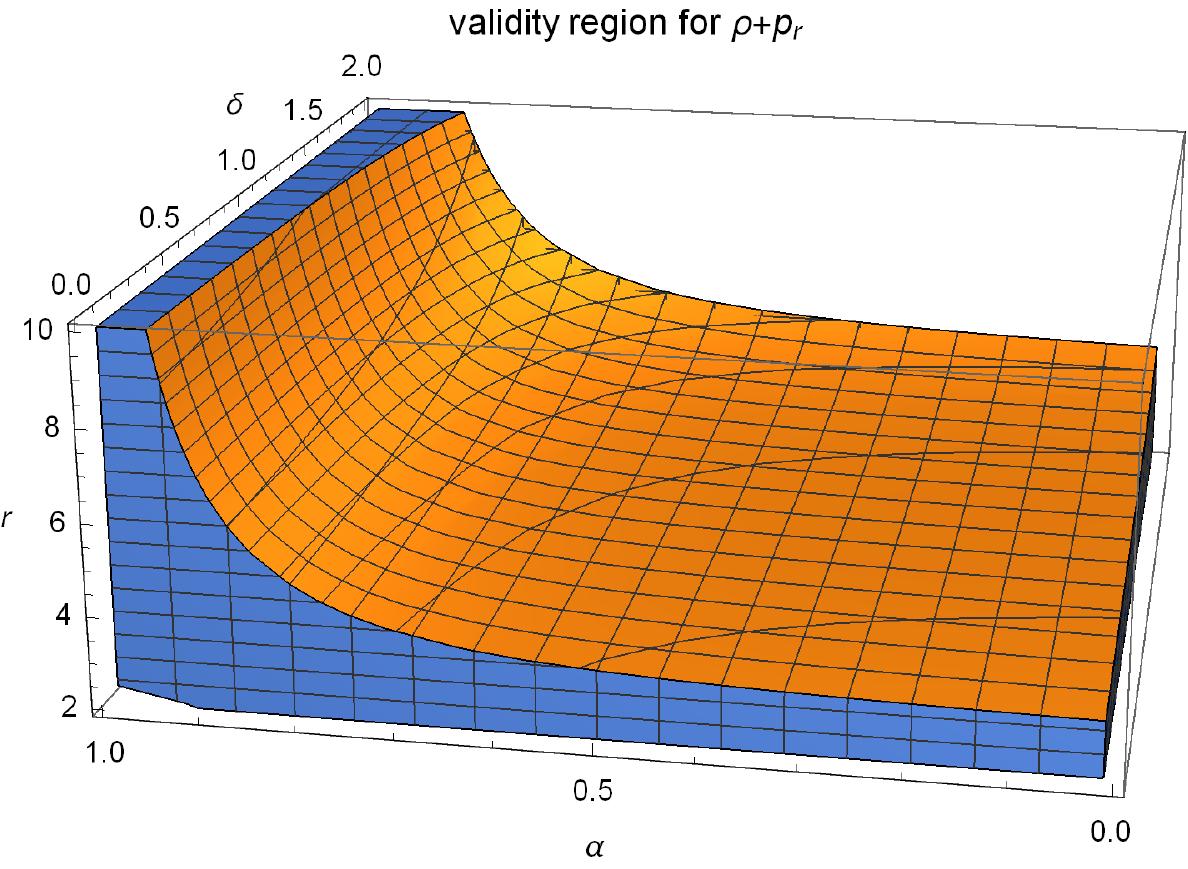, width=.45\linewidth,
height=2.5in}\caption{\label{Fig2} NEC ($\rho+p_r$) with $\gamma=-2$ (left) and $\gamma=6$ (right). for different values of $\alpha$ and $\delta$.}
\end{figure}
\begin{figure}
\centering \epsfig{file=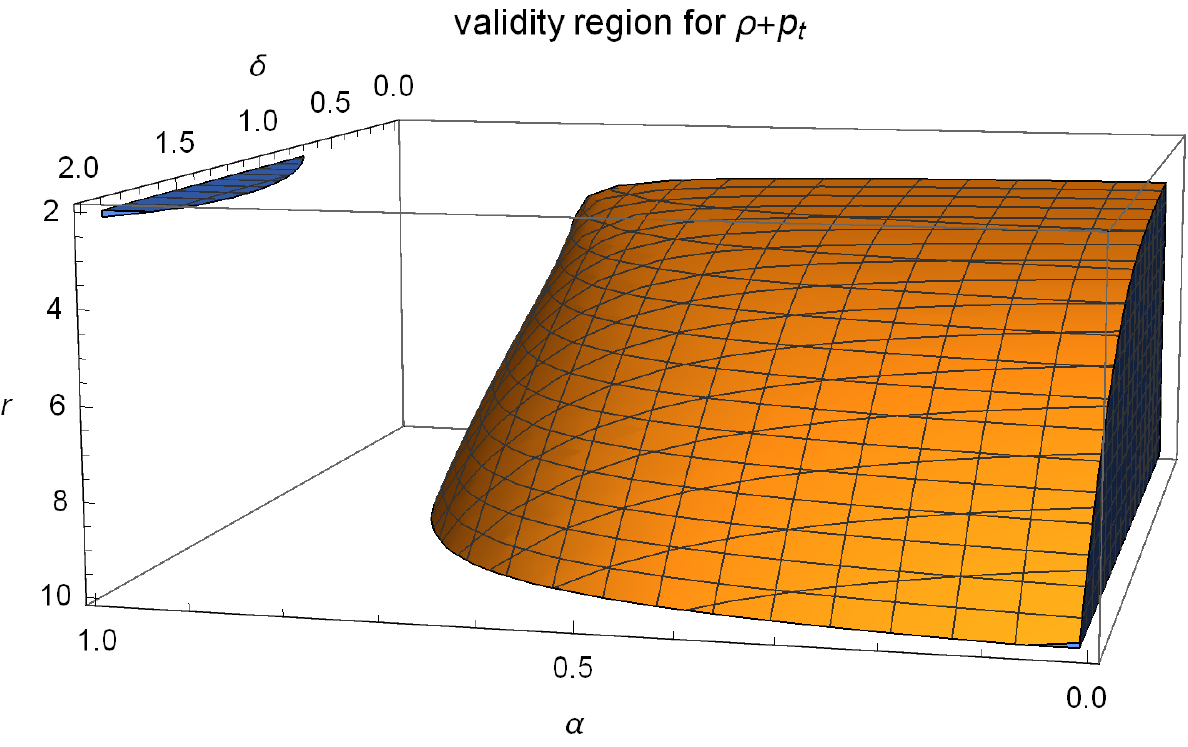, width=.45\linewidth,
height=2.5in}\epsfig{file=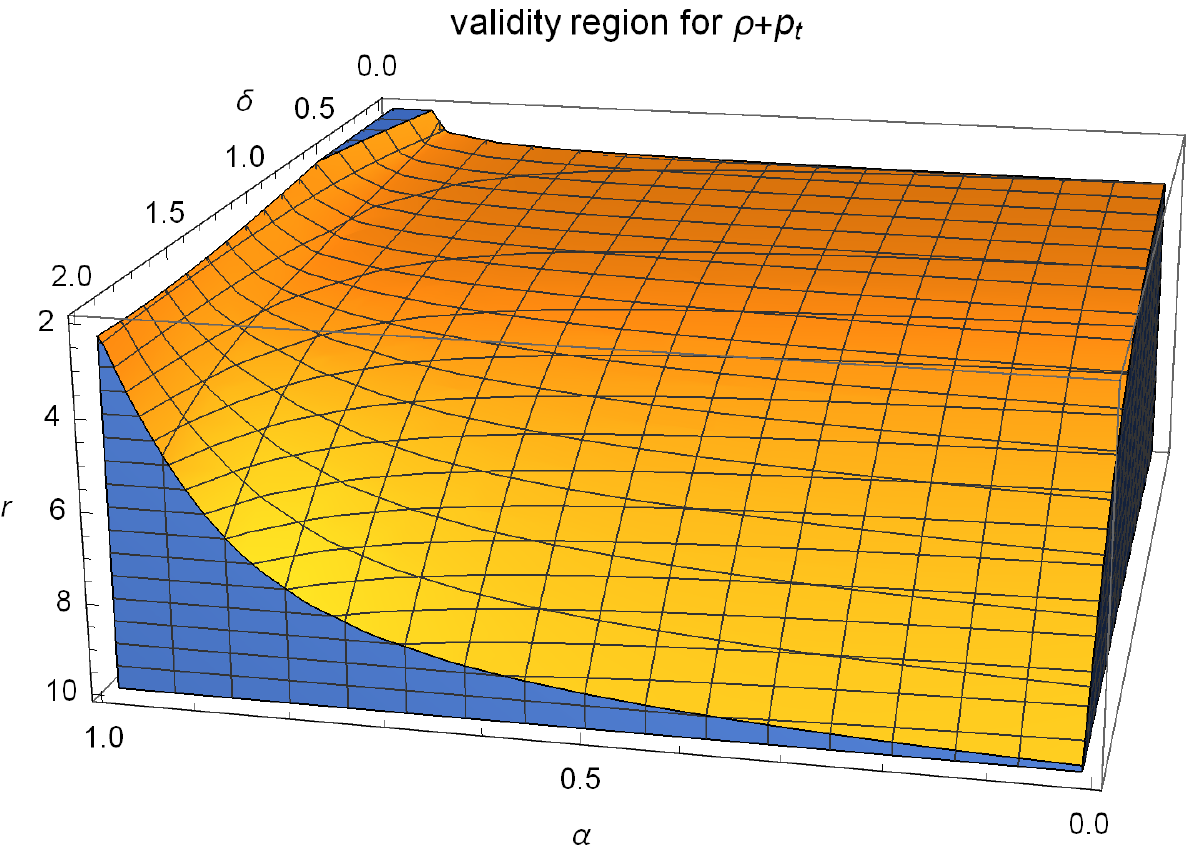, width=.45\linewidth,
height=2.5in}\caption{\label{Fig4} NEC ($\rho+p_t$) with $\gamma=-2$ (left) and $\gamma=6$ (right) for different values of $\alpha$ and $\delta$.}
\end{figure}
\begin{figure}
\centering \epsfig{file=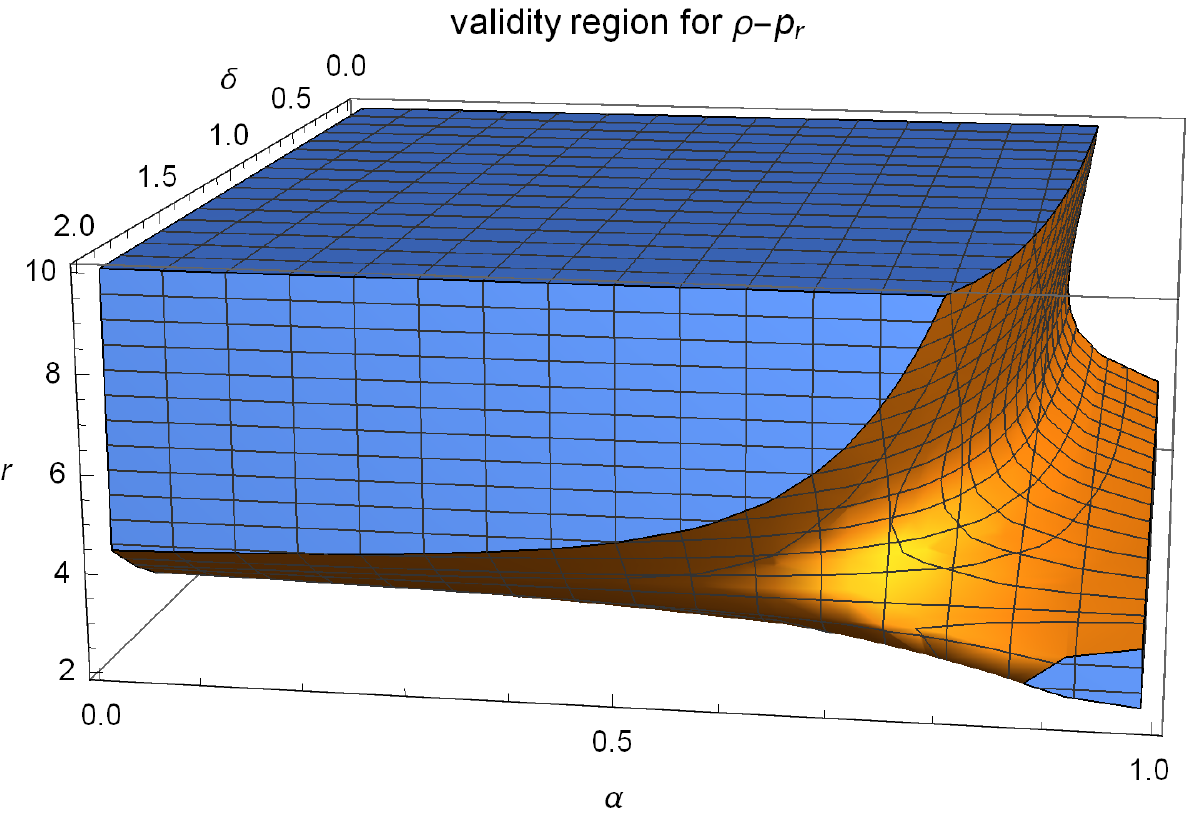, width=.45\linewidth,
height=2.5in}\epsfig{file=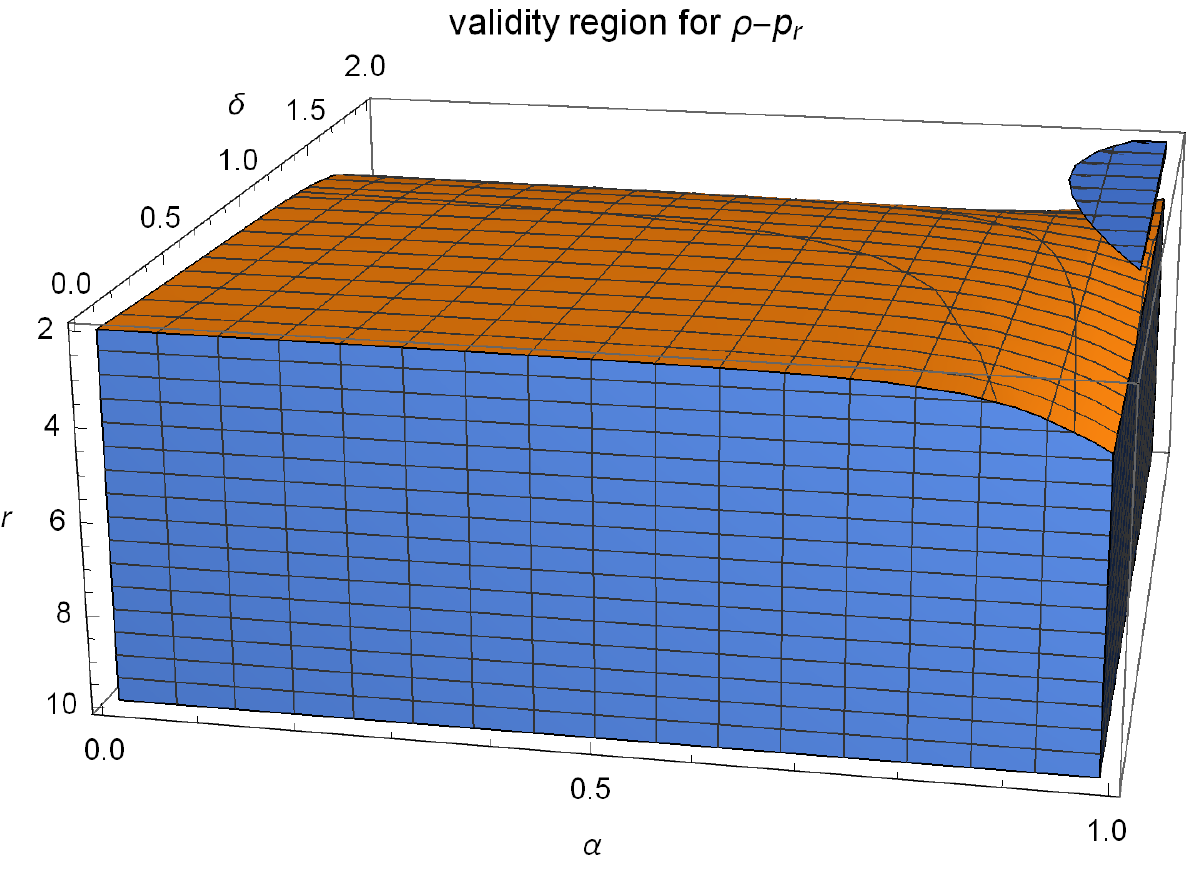, width=.45\linewidth,
height=2.5in}\caption{\label{Fig3} DEC ($\rho-p_r$) with $\gamma=-2$ (left) and $\gamma=6$ (right) for different values of $\alpha$ and $\delta$.}
\end{figure}
\begin{figure}
\centering \epsfig{file=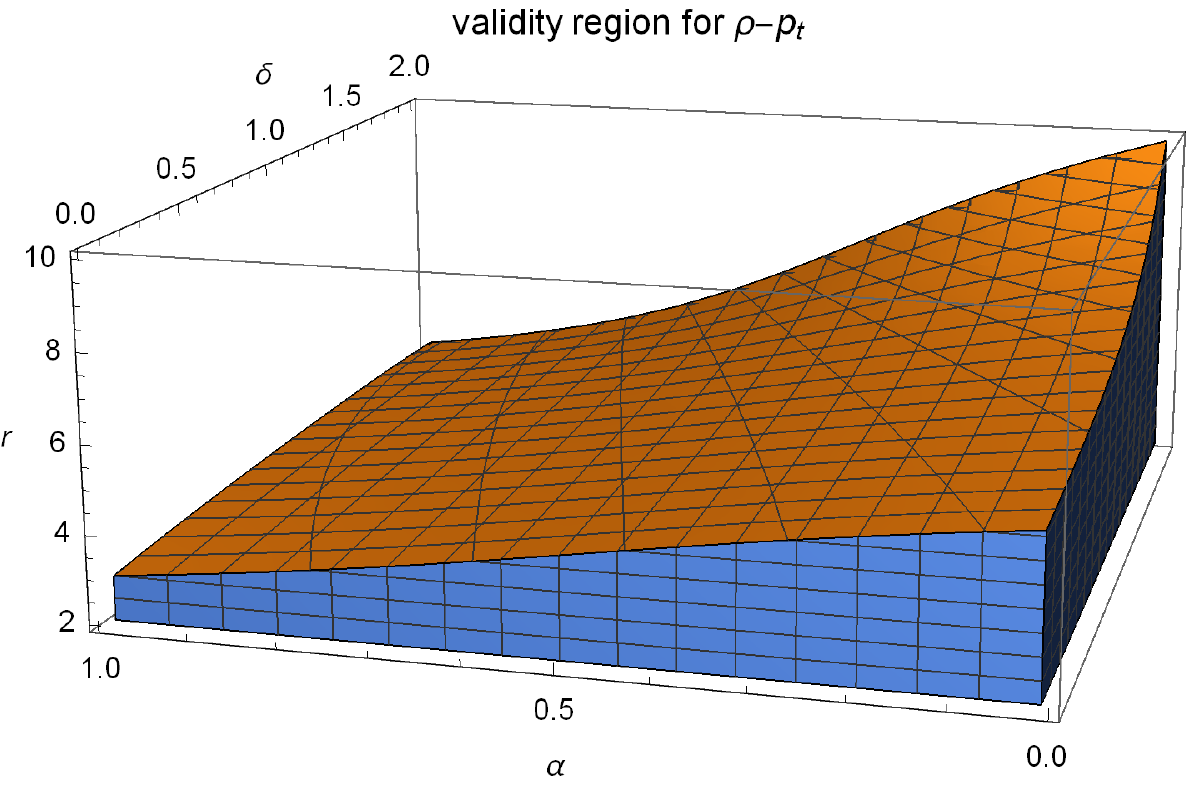, width=.45\linewidth,
height=2.5in}\epsfig{file=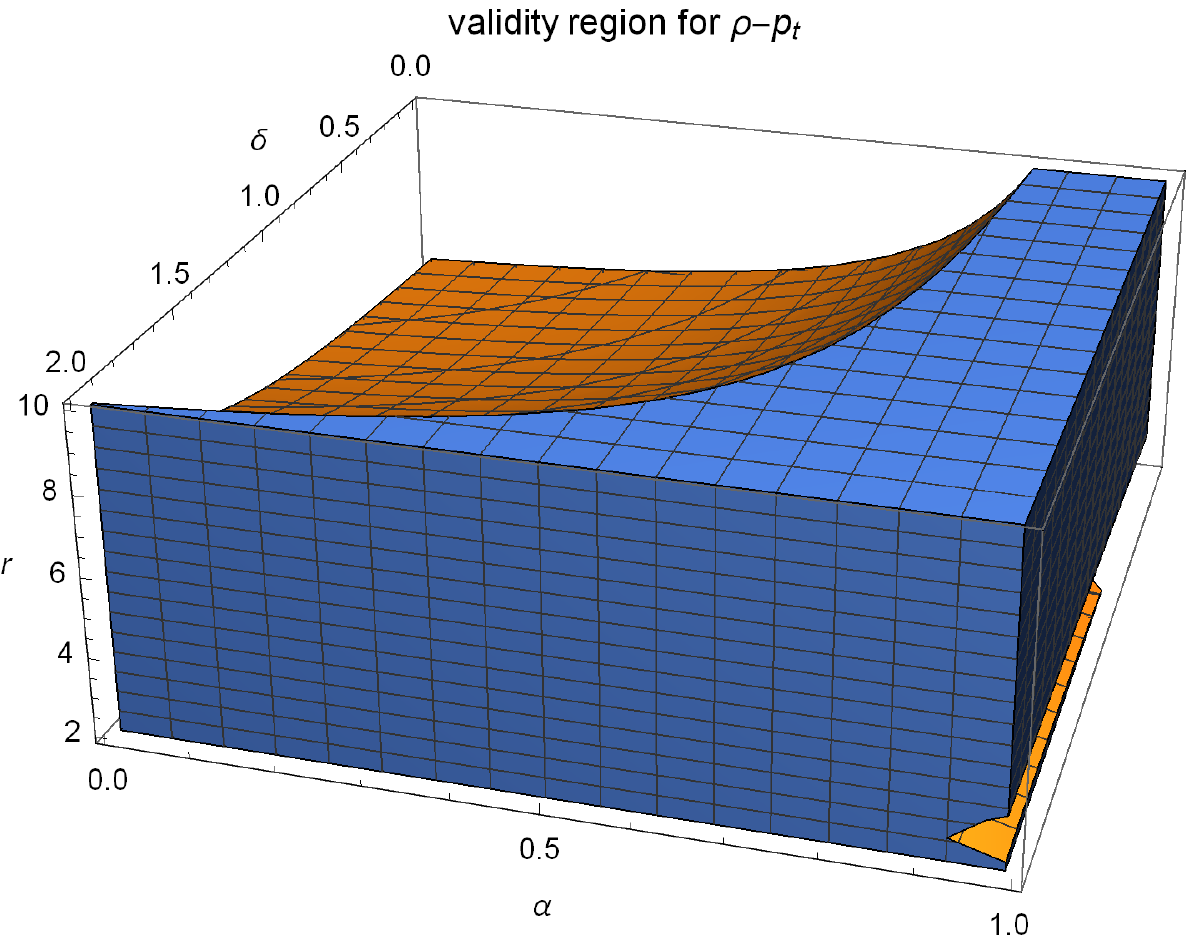, width=.45\linewidth,
height=2.5in}\caption{\label{Fig5} DEC ($\rho-p_t$) with $\gamma=-2$ (left) and $\gamma=6$ (right) for different values of $\alpha$ and $\delta$.}
\end{figure}
\begin{figure}
\centering \epsfig{file=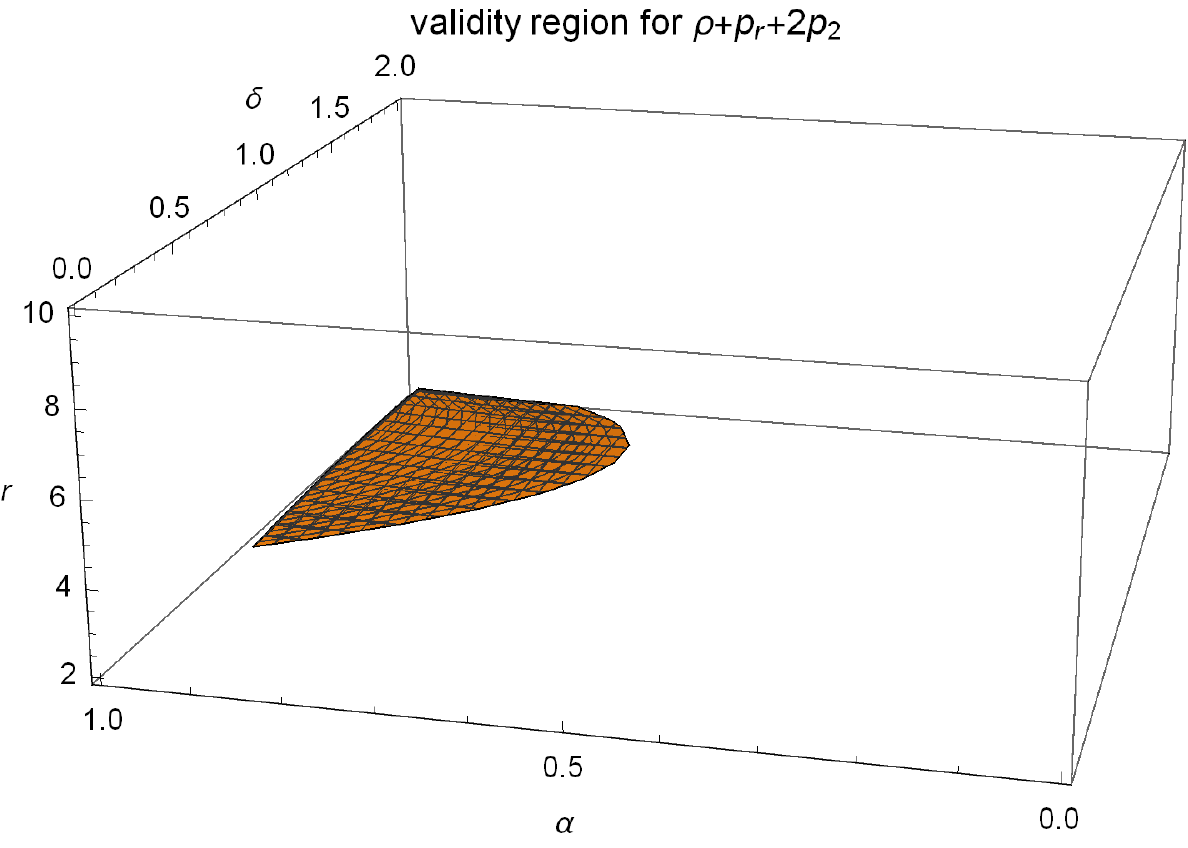, width=.45\linewidth,
height=2.5in}\epsfig{file=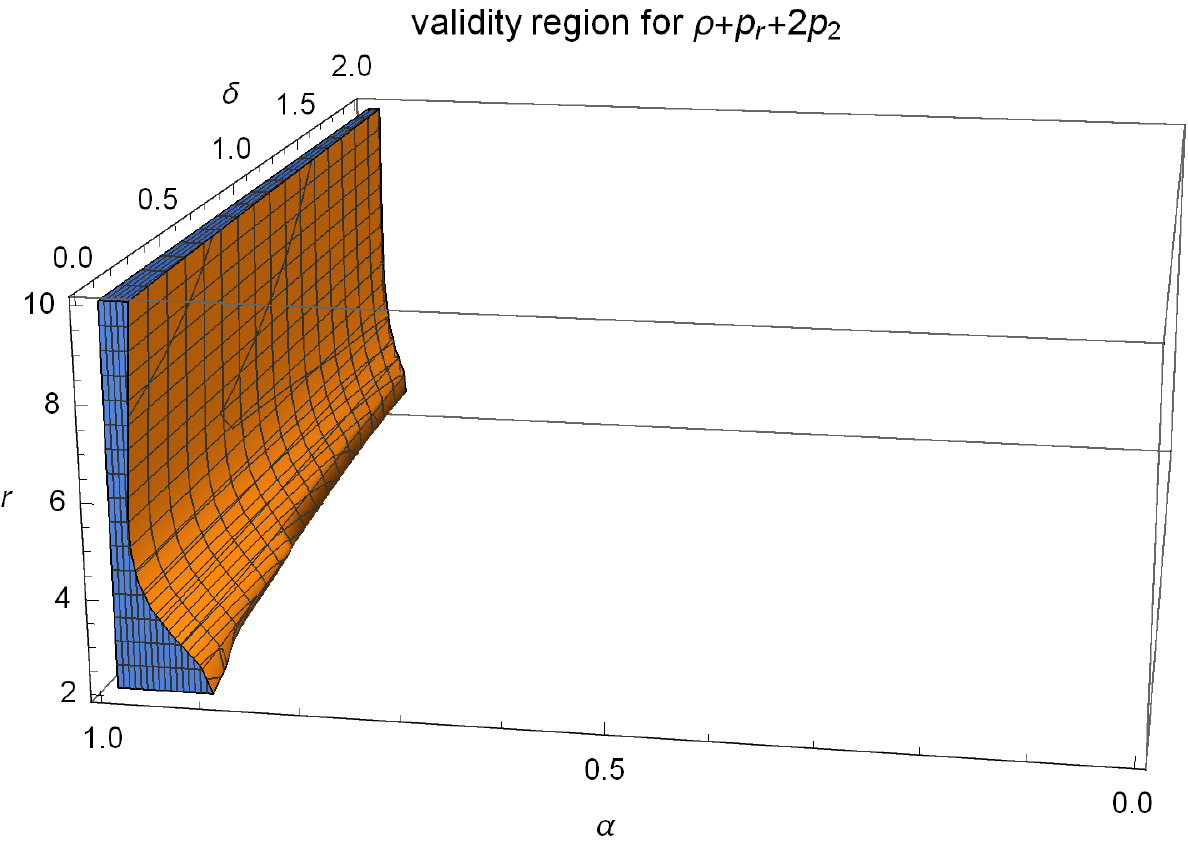, width=.45\linewidth,
height=2.5in}\caption{\label{Fig6} SEC ($\rho+p_r +2p_t$) with $\gamma=-2$ (left) and $\gamma=6$ (right) for different values of $\alpha$ and $\delta$.}
\end{figure}

\subsection{The $f(Q)=Q-\alpha \gamma \tanh \left(\frac{Q}{\gamma }\right)$ case}

Herein, we take 

\begin{equation}
\label{35}
f(Q)=Q-\alpha \gamma \tanh \left(\frac{Q}{\gamma }\right),
\end{equation}
where $0<\alpha<1$ and $\gamma$ is the model free parameter. By plugging Eq. (\ref{7}),  Eq. (\ref{8}), and Eq. (\ref{35}) into Eqs. (\ref{24}-\ref{26}), we obtain:
\begin{eqnarray}
\rho&=&\frac{1}{2} \left(\alpha  \gamma  \tanh \left(\psi _1\right)-\frac{1}{r^4}\left(\frac{16 \alpha  \psi _2 (r-b(r)) \tanh \left(\psi _1\right) \text{sech}^2\left(\psi _1\right)}{\gamma  r^3}-2 (r-b(r)) (4 \chi +r)\right)+\psi _3\right),\label{36}\\
p_{r}&=&\frac{b(r) (4 \chi +r) \text{sech}^2\left(\psi _1\right) \left(-4 \alpha +\cosh \left(\frac{4 (r-b(r)) (4 \chi +r)}{\gamma  r^4}\right)+1\right)+r \psi _4}{2 r^4},\label{37}\\
p_{t}&=&\frac{1}{2} \left(-\alpha  \gamma  \tanh \left(\psi _1\right)+\frac{1}{r^4}\left(2 (r-b(r)) (4 \chi +r)+\frac{8 \alpha  \psi _6}{\gamma  r^4}-\frac{\psi _5 \left(\alpha  \text{sech}^2\left(\psi _1\right)-1\right)}{r}\right)\right),\label{38}
\end{eqnarray}
where
\begin{eqnarray*}
\psi _1&&=\frac{2 (r-b(r)) (4 \chi +r)}{\gamma  r^4},\\
\psi _2&&=r \left((4 \chi +r) b'(r)+2 (6 \chi +r)\right)-b(r) (16 \chi +3 r),\\
\psi _3&&=\frac{2 \left(\alpha  \text{sech}^2\left(\psi _1\right)-1\right) \left(r \left(r b'(r)-4 \chi -r\right)+b(r) (4 \chi +r)\right)}{r^4},\\
\psi _4&&=-8 \chi -\alpha  \gamma  r^3 \tanh \left(\psi _1\right)+2 \alpha  (8 \chi +r) \text{sech}^2\left(\psi _1\right),\\
\psi _5&&=r (2 \chi +r) \left(r b'(r)-2 \left(2 \chi +r\right)\right)+b(r) \left(8 \chi ^2+r^2+2 \chi  r\right),\\
\psi _6&&=\psi _2 (r-b(r)) (2 \chi +r) \tanh \left(\psi _1\right) \text{sech}^2\left(\psi _1\right).
\end{eqnarray*}

We plot in Figures 7-12 the energy conditions for such material solutions.

\begin{figure}
\centering \epsfig{file=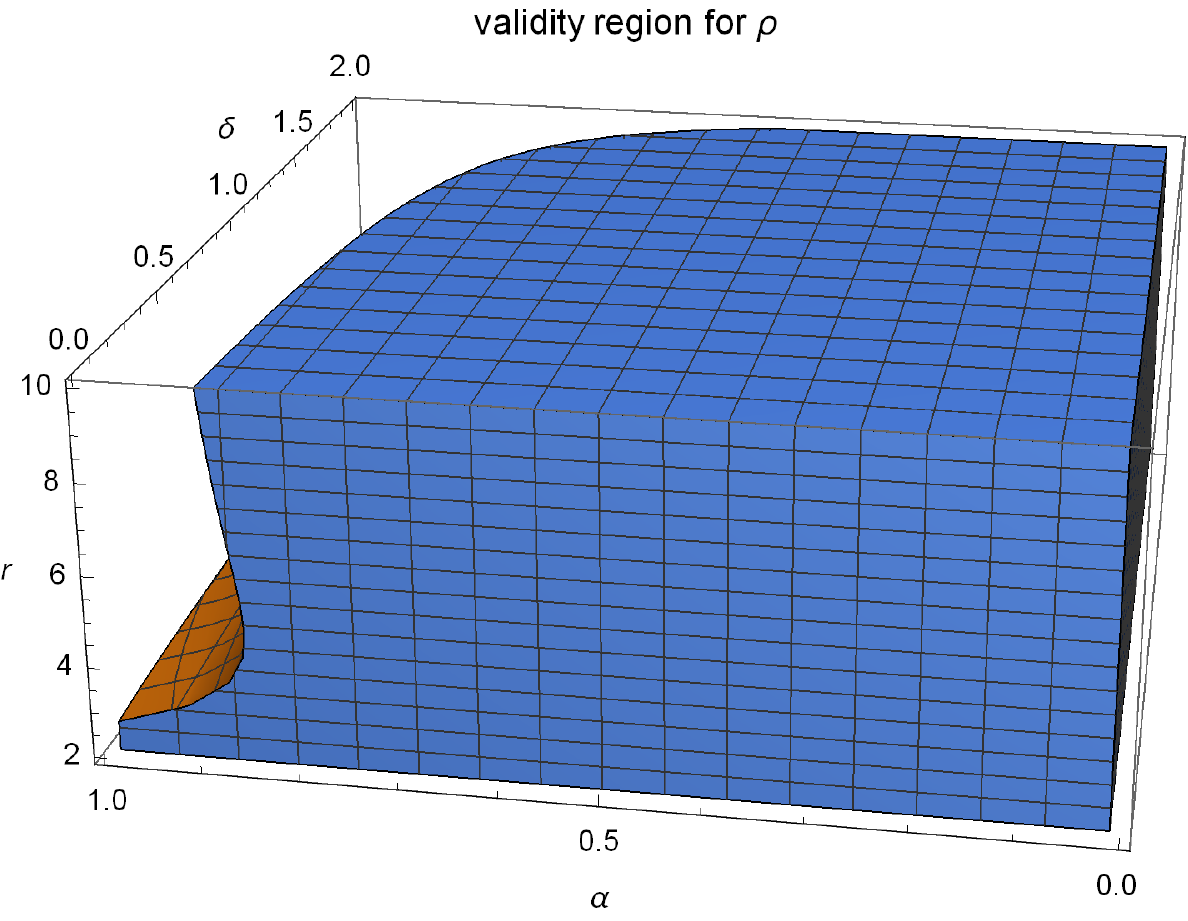, width=.45\linewidth,
height=2.5in}\epsfig{file=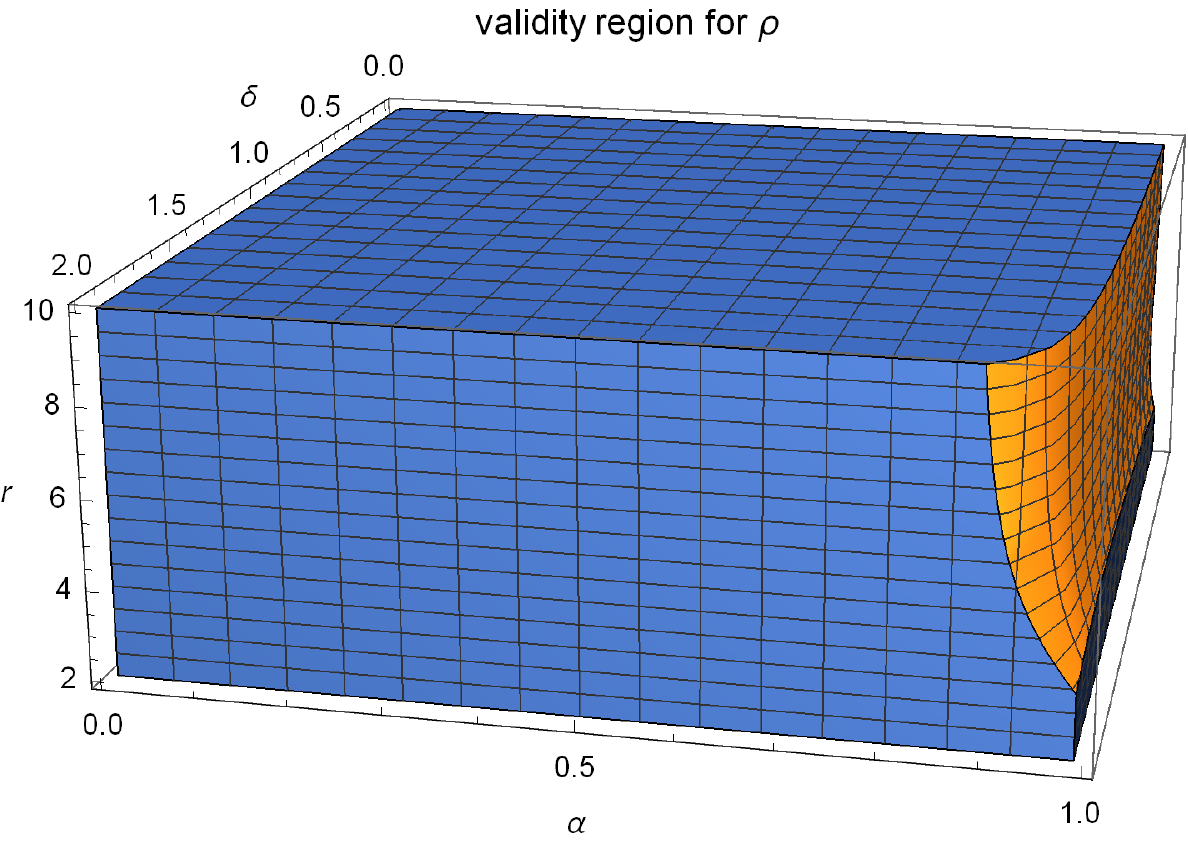, width=.45\linewidth,
height=2.5in}\caption{\label{Fig7} Energy density $\rho(r)$ with $\gamma=-2$ (left) and $\gamma=6$ (right) for different values of $\alpha$ and $\delta$.}
\end{figure}
\begin{figure}
\centering \epsfig{file=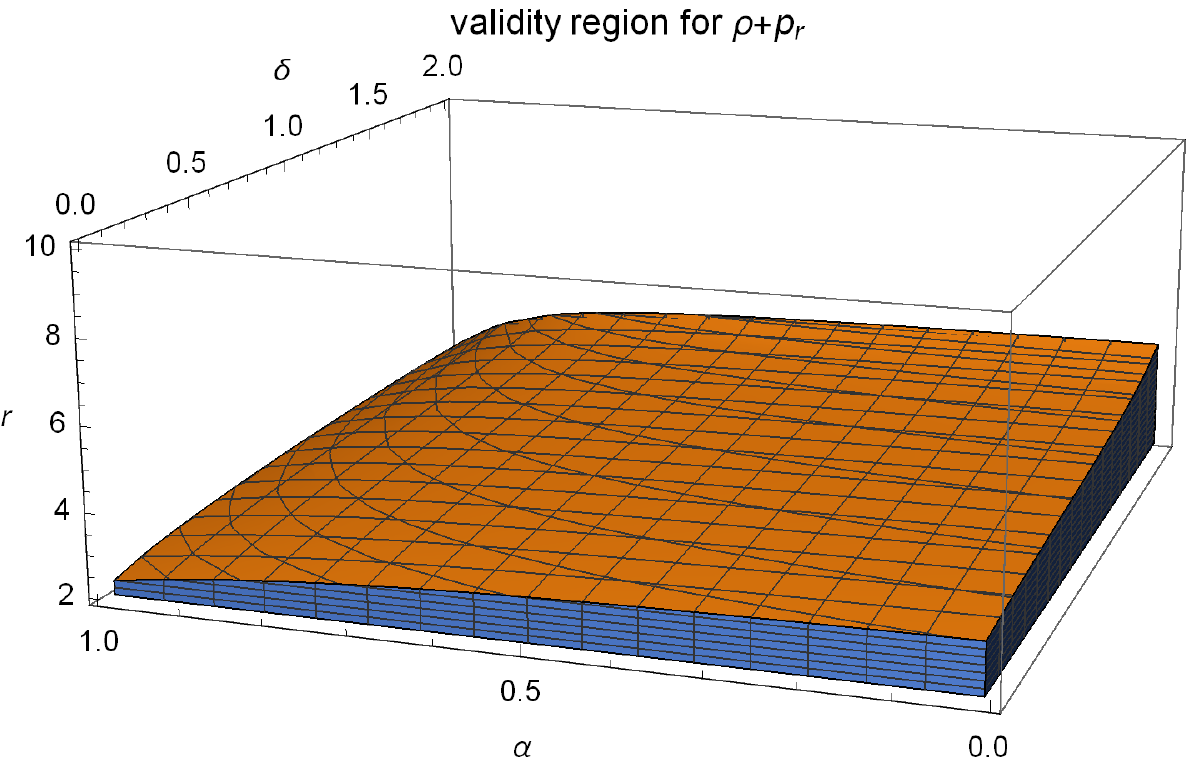, width=.45\linewidth,
height=2.5in}\epsfig{file=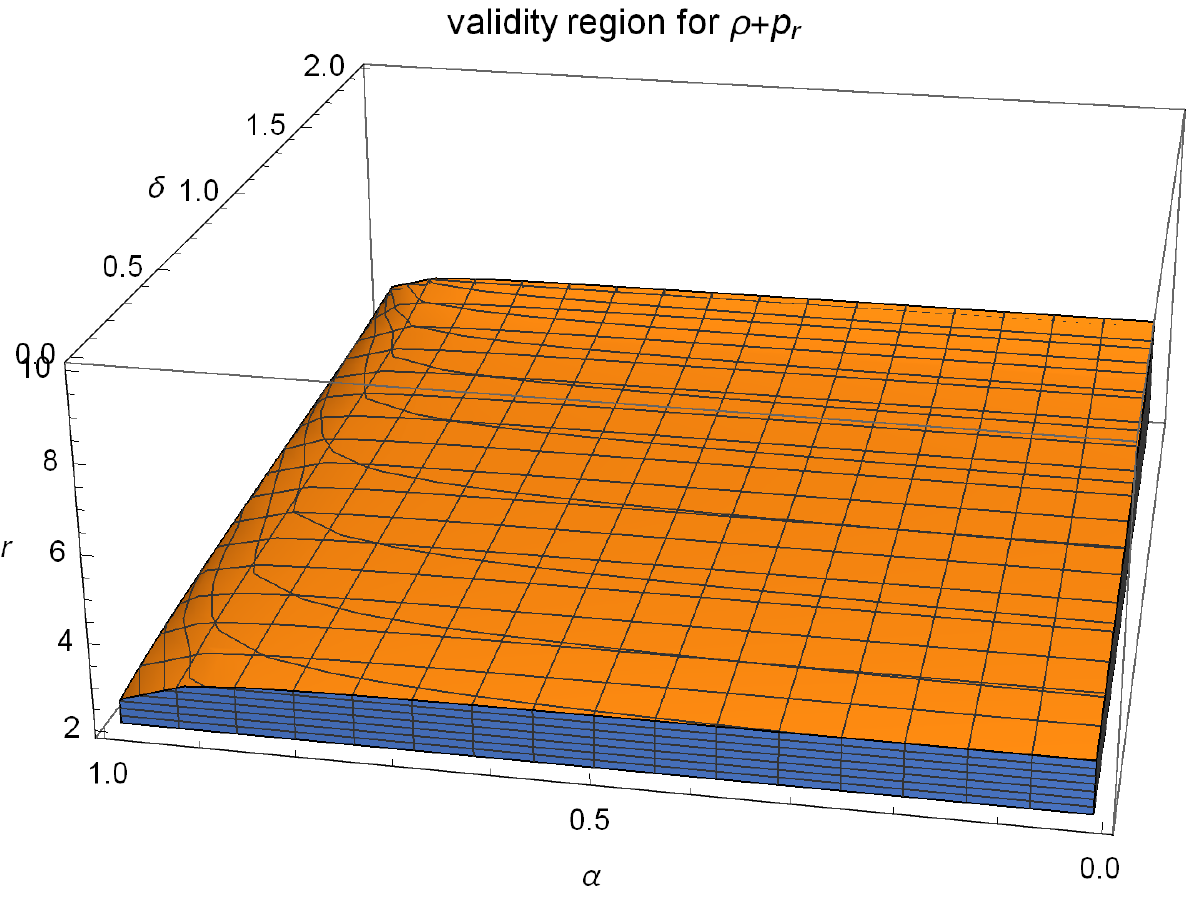, width=.45\linewidth,
height=2.5in}\caption{\label{Fig8} NEC ($\rho+p_r$) with $\gamma=-2$ (left) and $\gamma=6$ (right) for different values of $\alpha$ and $\delta$.}
\end{figure}
\begin{figure}
\centering \epsfig{file=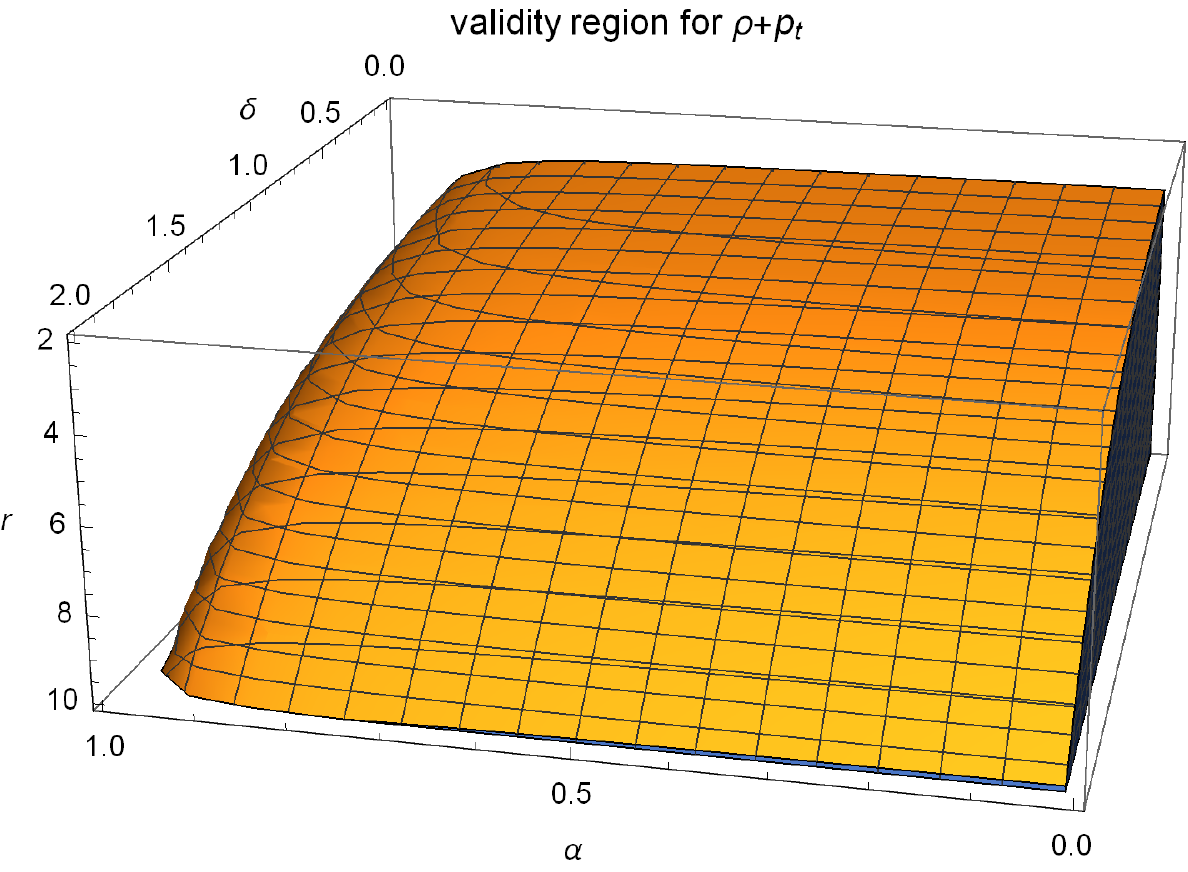, width=.45\linewidth,
height=2.5in}\epsfig{file=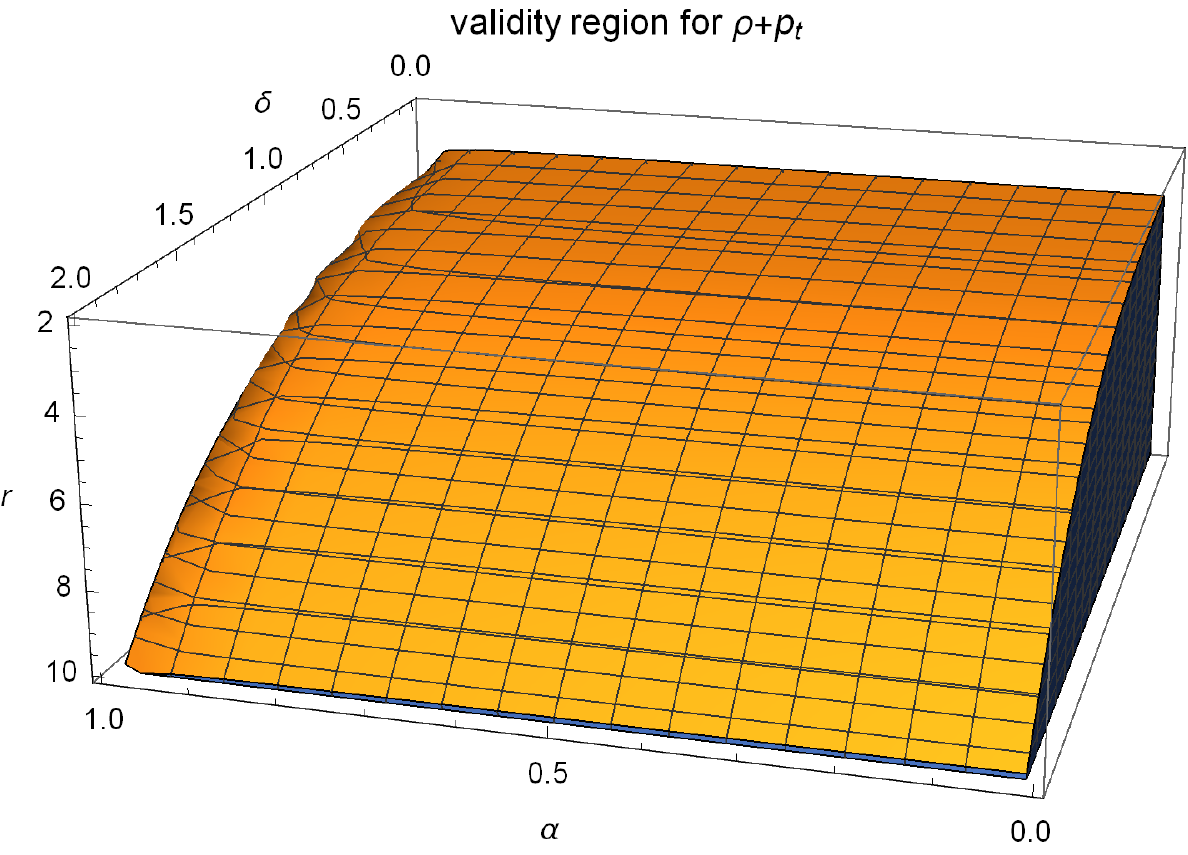, width=.45\linewidth,
height=2.5in}\caption{\label{Fig10} NEC ($\rho+p_t$) with $\gamma=-2$ (left) and $\gamma=6$ (right) for different values of $\alpha$ and $\delta$.}
\end{figure}
\begin{figure}
\centering \epsfig{file=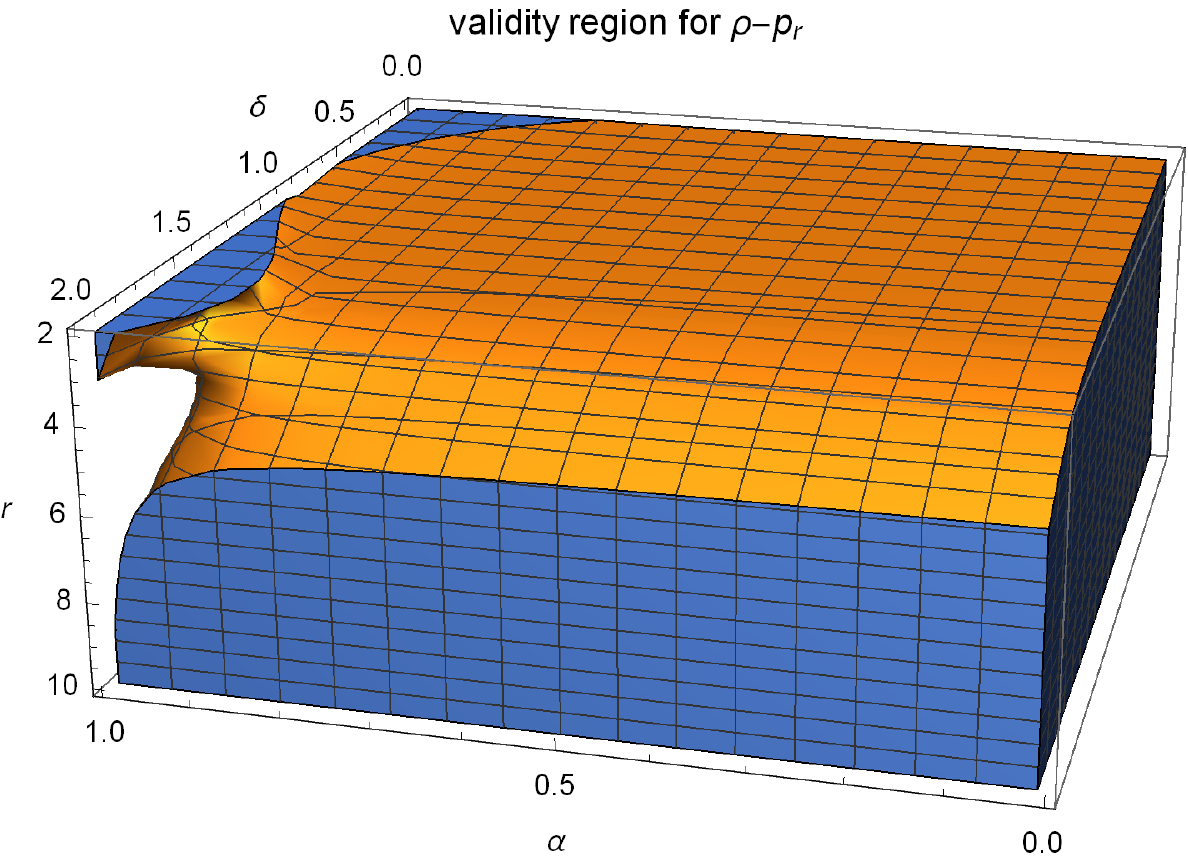, width=.45\linewidth,
height=2.5in}\epsfig{file=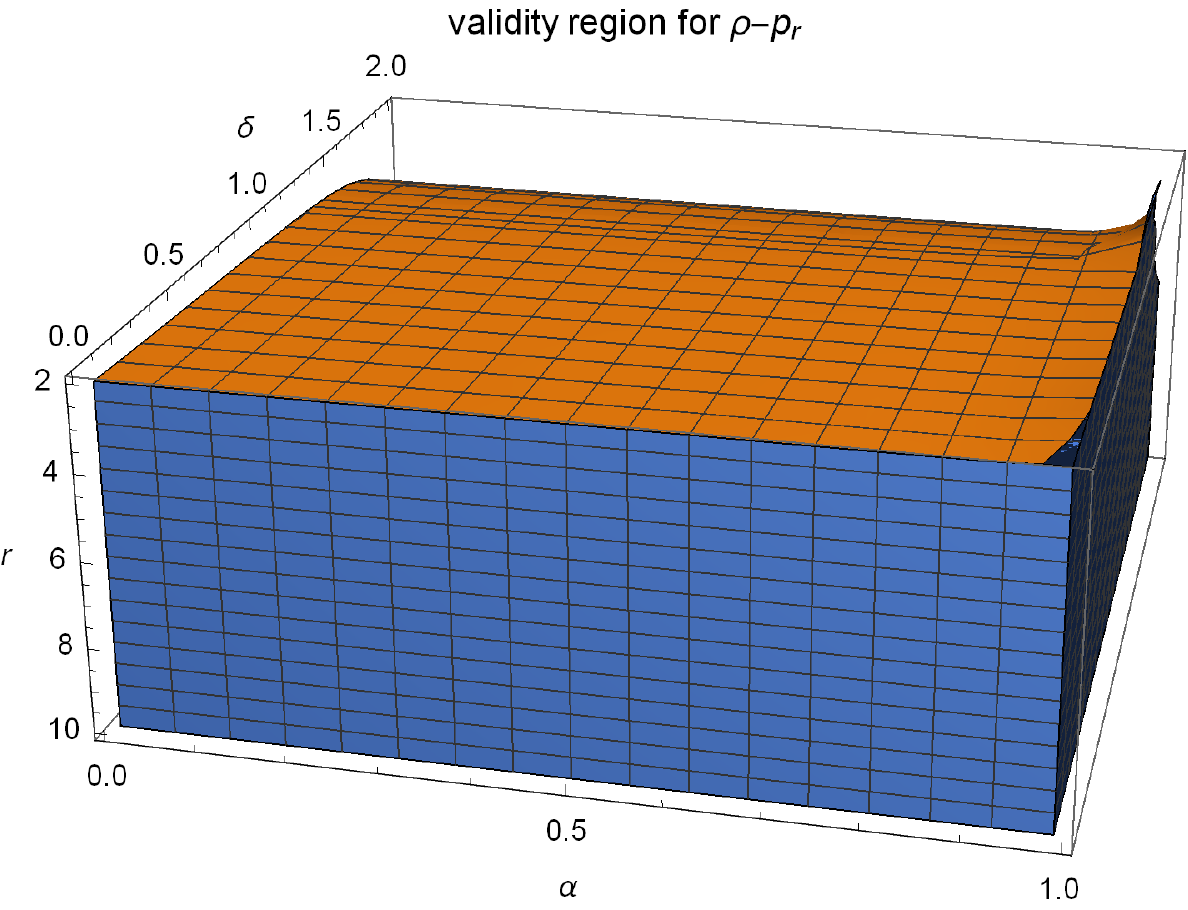, width=.45\linewidth,
height=2.5in}\caption{\label{Fig9} DEC ($\rho-p_r$) with $\gamma=-2$ (left) and $\gamma=6$ (right) for different values of $\alpha$ and $\delta$.}
\end{figure}
\begin{figure}
\centering \epsfig{file=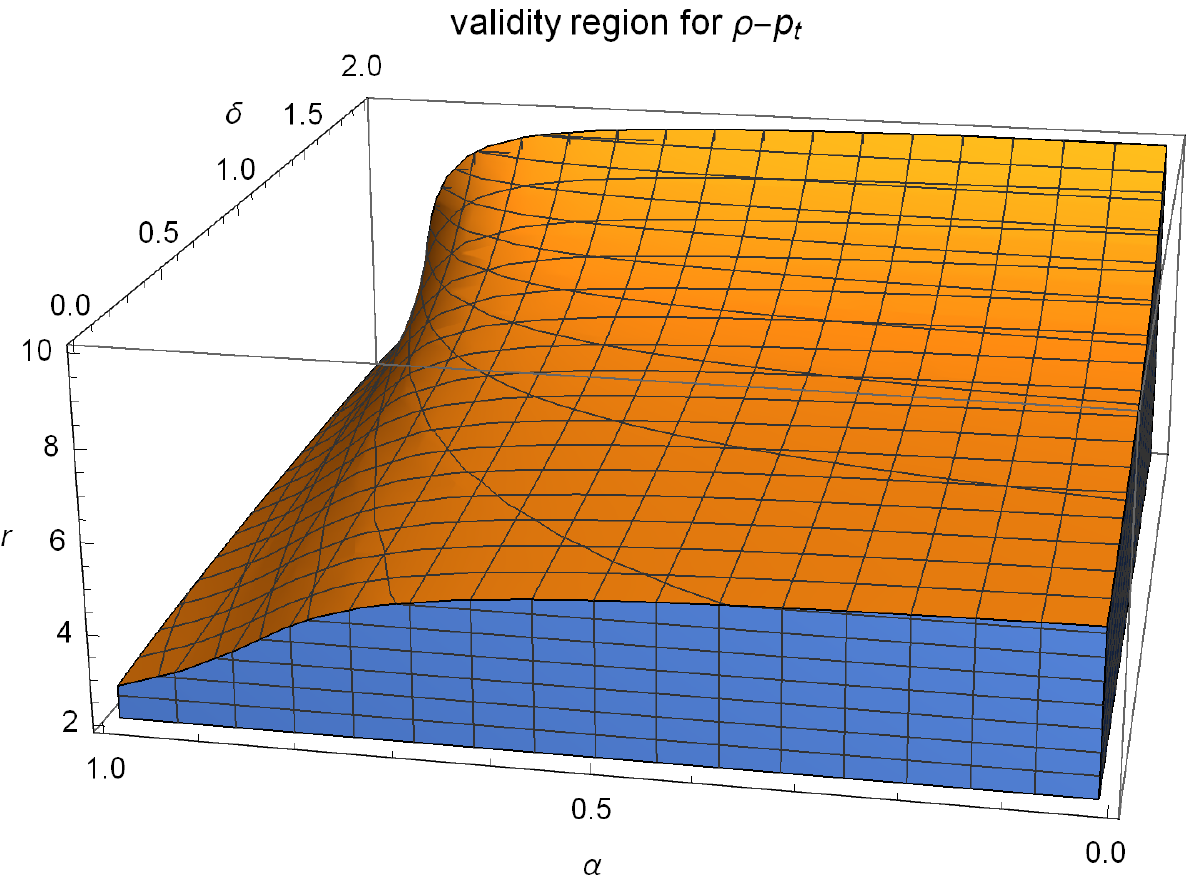, width=.45\linewidth,
height=2.5in}\epsfig{file=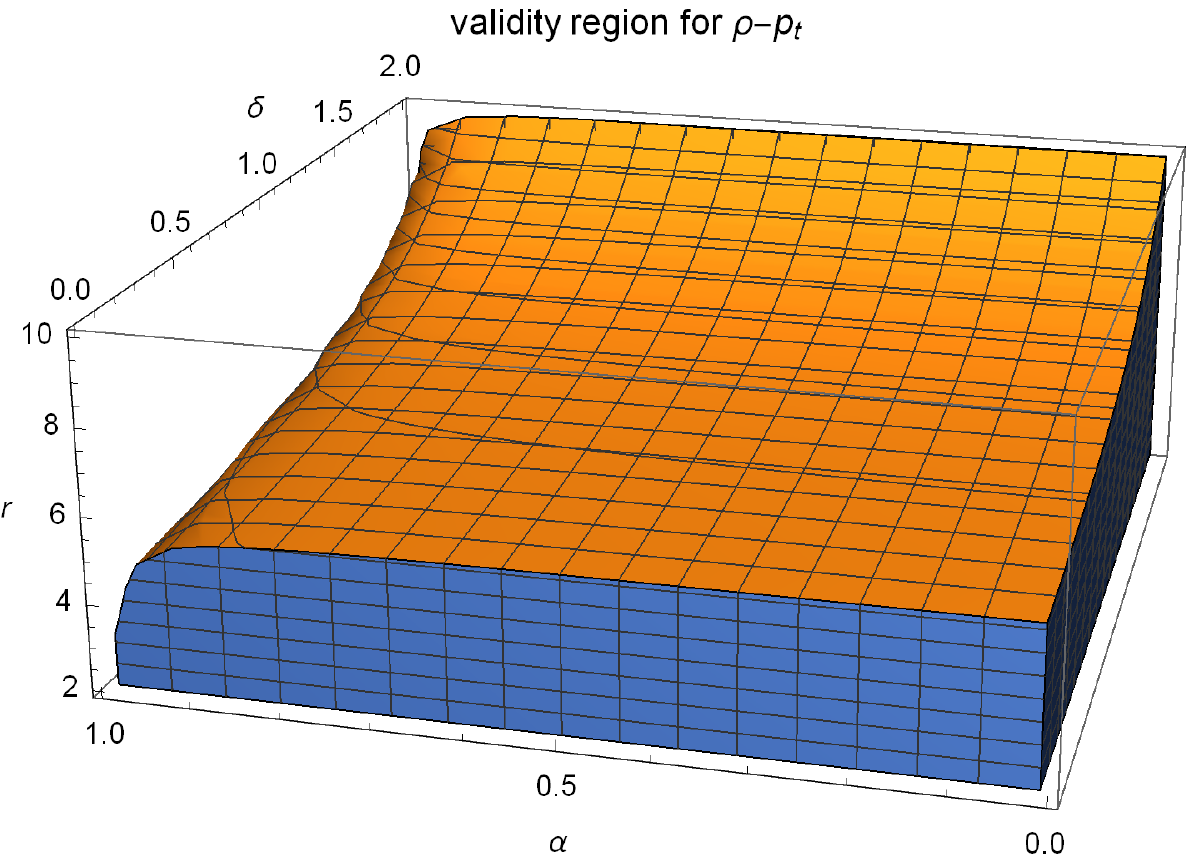, width=.45\linewidth,
height=2.5in}\caption{\label{Fig11} DEC ($\rho-p_t$) with $\gamma=-2$ (left) and $\gamma=6$ (right) for different values of $\alpha$ and $\delta$.}
\end{figure}
\begin{figure}
\centering \epsfig{file=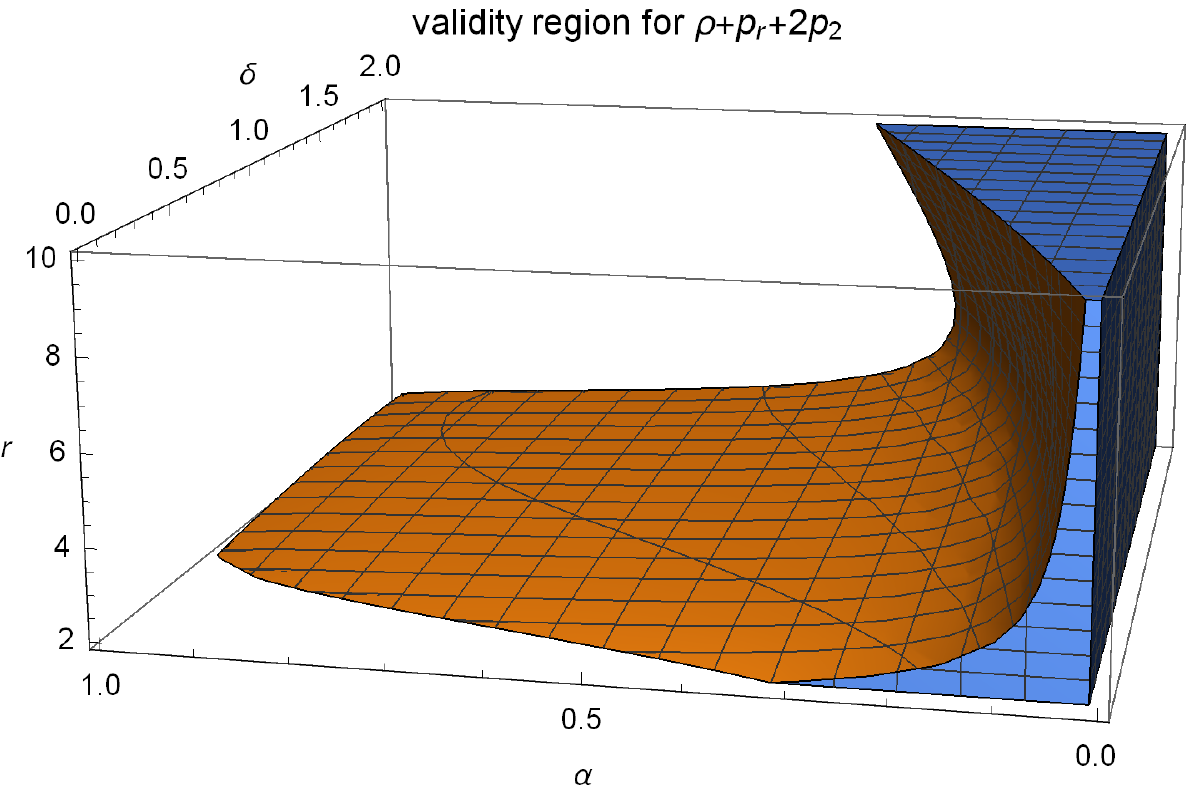, width=.45\linewidth,
height=2.5in}\epsfig{file=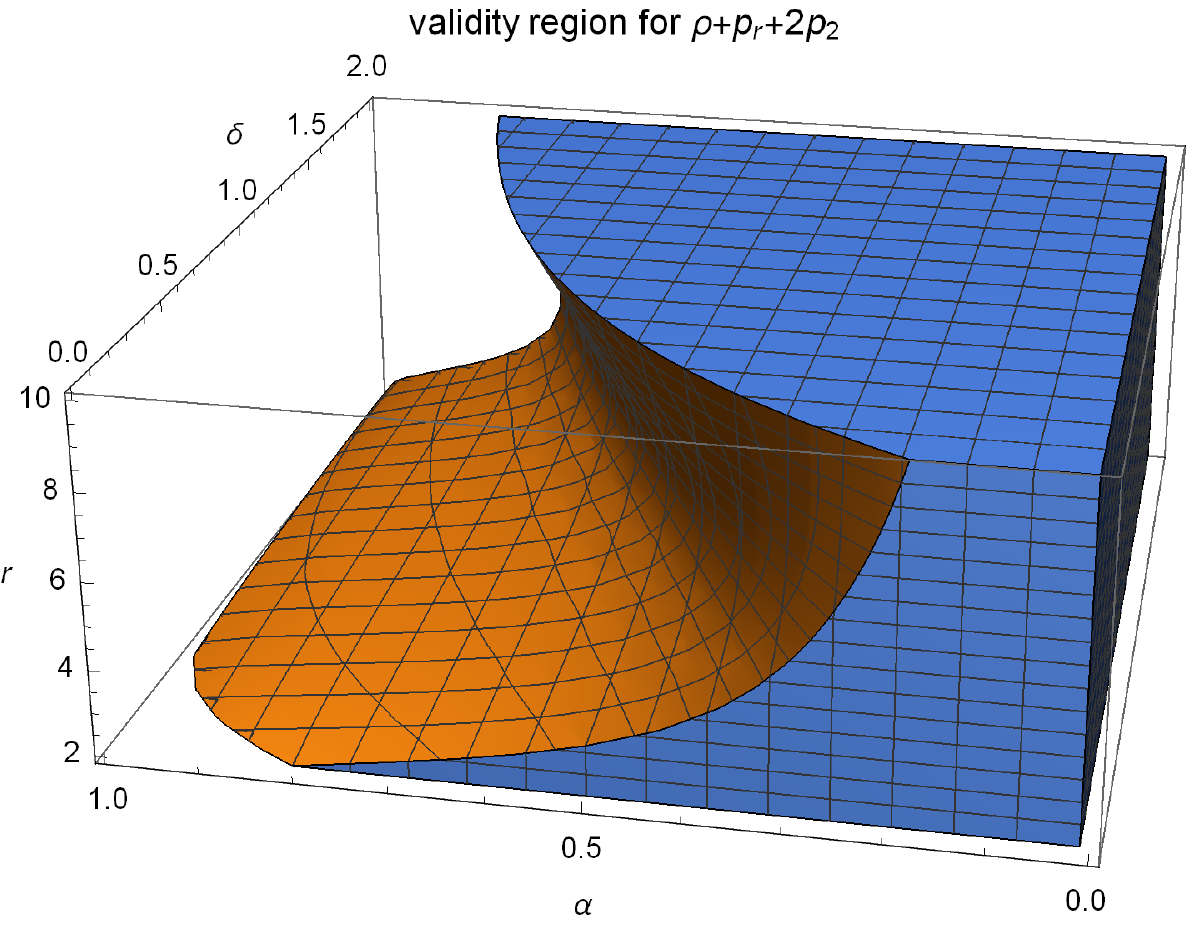, width=.45\linewidth,
height=2.5in}\caption{\label{Fig12} SEC ($\rho+p_r+2p_t$) with $\gamma=-2$ (left) and $\gamma=6$ (right) for different values of $\alpha$ and $\delta$.}
\end{figure}

\section{Conclusion}

Wormholes are solutions of Einstein's GR field equations that have not yet been observed, although attempts to do so are constantly proposed in the literature (besides References \cite{rahaman/2014,ovgun/2016,safonova/2002,bambi/2021}, presented in Introduction Section, one can also check \cite{dai/2019,cramer/1995,wang/2020,ohgami/2015,torres/1998,shaikh/2018,nandi/2006,jusufi/2018,damour/2007}). 

Although the wormhole concept predate Morris and Thorne article \cite{Morris}, in \cite{Morris} the concept of {\it traversable} wormholes was first introduced. According to Morris and Thorne, to be traversable, a wormhole should obey a series of properties both on the geometrical and material sectors. While the metric conditions were presented and respected in Section II, GR wormhole material content must be exotic, in the sense that it should disrespect some energy conditions and even present negative mass.

Here we have obtained wormhole material solutions for two $f(Q)$ cases and constructed the energy conditions for each of them. In the first case, all the energy conditions can be satisfied at least for a small range of the model free parameters and for both values assumed for $\gamma$, as it is the case of SEC in Fig.6, which points to the necessity of SEC violating matter as one steps away from the wormhole throat. This is not a very critical situation as a classical free minimally-coupled massive scalar field $\Phi$ with mass $m$ described by 

\begin{equation}
	\mathcal{L}_m=\frac{1}{2}[(\nabla\Phi)^2+(m\Phi)^2]
\end{equation}
can easily violate SEC \cite{hawking/1973,visser/1995}.

For the second case, once again it is possible to respect the energy conditions, with the NEC ($\rho+p_r$) being respected for small values of $r$ for both values of $\gamma$ while SEC respectability is more easily attained than the first $f(Q)$ case, specially for smaller values of $\alpha$, as one can check Fig.12 left panel.\\

{\bf Acknowledgements}\\

PHRSM thanks CAPES for financial support. We are very much grateful to the honorable referee and the
editor for the illuminating suggestions that have significantly improved our
work in terms of research quality and presentation.

\end{document}